\documentclass{mn2e} 
\usepackage{color}
\usepackage{dcolumn}
\usepackage{bm}
\usepackage{hyperref}
\usepackage{epsfig}

\def\reference{\par\noindent\hangindent\parindent}

 at 13 truept

\begin{document}

\twocolumn

\title[GAMA Panchromatic Data Release]{Galaxy And Mass Assembly (GAMA): Panchromatic Data Release (far-UV---far-IR) and the low-z energy budget}

\author[Driver et al.] 
{Simon~P.~Driver,$^{1,2}$\thanks{SUPA, Scottish Universities Physics Alliance}\thanks{e-mail:Simon.Driver@uwa.edu.au}, 
Angus~H. Wright$^1$, 
Stephen~K. Andrews$^1$,
Luke~J. Davies$^1$,
\newauthor
Prajwal~R. Kafle$^1$,
Rebecca Lange$^1$,
Amanda~J. Moffett$^1$,
Elizabeth Mannering$^1$,
\newauthor
Aaron~S.~G. Robotham$^1$,
Kevin Vinsen$^1$,
Mehmet Alpaslan$^{3}$,
Ellen Andrae$^{4}$, 
\newauthor
Ivan~K. Baldry$^5$,
Amanda~E. Bauer$^{6}$,
Steven~P. Bamford$^7$,
Joss Bland-Hawthorn$^8$,
\newauthor
Nathan Bourne$^{9}$,
Sarah Brough$^6$,
Michael~J.~I. Brown$^{10}$,
Michelle~E. Cluver$^{11}$,
\newauthor
Scott Croom$^{8}$,
Matthew Colless$^{12}$,
Christopher~J. Conselice$^{7}$,
Elisabete da Cunha$^{13}$,
\newauthor
Roberto De Propris$^{14}$,
Michael Drinkwater$^{15}$,
Loretta Dunne$^{9,16}$,
Steve Eales$^{16}$,
\newauthor
Alastair Edge$^{17}$,
Carlos Frenk$^{18}$,
Alister~W. Graham$^{13}$,
Meiert Grootes$^{4}$,
\newauthor
Benne~W. Holwerda$^{19}$,
Andrew~M. Hopkins$^6$,
Edo Ibar$^{20}$,
Eelco van Kampen$^{21}$
\newauthor
Lee~S. Kelvin$^{5}$,
Tom Jarrett$^{22}$, 
D. Heath Jones$^{23}$,
Maritza~A. Lara-Lopez$^{24}$,
\newauthor
Jochen Liske$^{25}$,
Angel~R. Lopez-Sanchez$^{6}$,
Jon Loveday$^{26}$,
Steve~J. Maddox$^{9,16}$, 
\newauthor
Barry Madore$^{27}$,
Smriti Mahajan$^{28}$,
Martin Meyer$^1$,
Peder Norberg$^{17,18}$,
\newauthor
Samantha~J. Penny$^{29}$,
Steven Phillipps$^{30}$,
Cristina Popescu$^{31}$
Richard~J. Tuffs$^{4}$,
\newauthor
John~A. Peacock$^{9}$,
Kevin~A. Pimbblet$^{10,32}$,
Matthew Prescott$^{11}$,
Kate Rowlands$^{2}$,
\newauthor
Anne~E. Sansom$^{31}$,
Mark Seibert$^{27}$, 
Matthew~W.L. Smith$^{15}$,
Will~J. Sutherland$^{33}$, 
\newauthor
Edward~N. Taylor$^{34}$,
Elisabetta Valiante$^{16}$
J.~Antonio Vazquez-Mata$^{26}$,
Lingyu Wang$^{18,35}$,
\newauthor
Stephen M. Wilkins$^{26}$,
Richard Williams$^{5}$
\\
$^1$ICRAR\thanks{International Centre for Radio Astronomy Research}, The University of Western Australia, 35 Stirling Highway, Crawley, WA 6009, Australia\\
$^2$SUPA\thanks{Scottish Universities Physics Alliance}, School of Physics \& Astronomy, University of St Andrews, North Haugh, St Andrews, KY16 9SS, UK \\
$^3$ NASA Ames Research Centre, N232, Moffett Field, Mountain View, 94035 CA, United States \\
$^4$ Max Planck Institute for Nuclear Physics (MPIK), Saupfercheckweg 1, 69117 Heidelberg, Germany \\
$^5$ Astrophysics Research Institute, Liverpool John Moores University, IC2, Liverpool Science Park, 146 Brownlow Hill, Liverpool, L3 5RF, UK \\
$^6$ Australian Astronomical Observatory, PO Box 915, North Ryde, NSW 1670, Australia\\
$^7$ Centre for Astronomy and Particle Theory, University of Nottingham, University Park, Nottingham NG7 2RD, UK\\
$^8$ Sydney Institute for Astronomy, School of Physics, University of Sydney, NSW 2006, Australia\\
$^9$ SUPA, Institute for Astronomy, University of Edinburgh, Royal Observatory, Blackford Hill, Edinburgh EH9 3HJ, UK\\
$^{10}$ School of Physics, Monash University, Clayton, Victoria 3800, Australia\\
$^{11}$ Astrophysics Group, The University of Western Cape, Robert Sobukwe Road, Bellville 7530, South Africa \\
$^{12}$ Research School of Astronomy and Astrophysics, Australian National University, Canberra, ACT 2611, Australia \\
$^{13}$ Centre for Astrophysics and Supercomputing, Swinburne University of Technology, Hawthorn, Victoria 3122, Australia \\
$^{14}$ Finnish Centre for Astronomy with ESO, University of Turku, V\"ais\"al\"antie 20, Piikki\"o, 21500, Finland\\
$^{15}$ School of Mathematics and Physics, University of Queensland, Brisbane, QLD 4072, Australia \\
$^{16}$ School of Physics and Astronomy, Cardiff University, Queens Buildings, The Parade, Cardiff CF24 3AA, UK \\
$^{17}$ Centre for Extragalactic Astronomy, Department of Physics, Durham University, South Road, Durham, DH1 3LE, UK\\
$^{18}$ Institute for Computational Cosmology, Department of Physics, Durham University, South Road, Durham, DH1 3LE, UK\\
$^{19}$ Leiden Observatory, University of Leiden, Niels Bohrweg 2, 2333 CA, Leiden, The Netherlands \\
$^{20}$ Instituto de F\'isica y Astronom\'ia, Universidad de Valpara\'iso, Avda. Gran Breta\~na 1111, Valpara\'iso, Chile \\
$^{21}$ European Southern Observatory, Karl-Schwarzschild-Str.~2, 85748 Garching, Germany\\
$^{22}$ Department of Astronomy, University of Cape Town, Private Bag X3, Rondebosch 7701, South Africa \\
$^{23}$ Department of Physics and Astronomy, Macquarie University, Sydney, NSW 2109, Australia \\
$^{24}$ Instituto de Astronom\'ia, Universidad Nacional Aut\'omana de M\'exico, A.P. 70-264, 04510 M\'exico, D.F., M\'exico\\
$^{25}$ Hamburger Sternwarte, Universit{\"a}t Hamburg, Gojenbergsweg 112, 21029 Hamburg, Germany \\
$^{26}$ Astronomy Centre, Department of Physics and Astronomy, University of Sussex, Falmer, Brighton BN1 9QH, UK\\
$^{27}$ Observatories of the Carnegie Institute for Science, 813 Santa Barbara Street, Pasadena, CA 91101, USA \\
$^{28}$ Indian Institute of Science Education and Research Mohali, Knowledge City, Sector 81, Manauli 140306, Punjab, India\\
$^{29}$ Institute of Cosmology and Gravitation, University of Portsmouth, Dennis Sciama Building, Burnaby Road, Portsmouth, PO1 3FX, UK\\
$^{30}$ Astrophysics Group, H.H. Wills Physics Laboratory, University of Bristol, Tyndall Avenue, Bristol BS8 1TL, UK\\
$^{31}$ Jeremiah Horrocks Institute, University of Central Lancashire, Preston, Lancashire, PR1 2HE, UK \\
$^{32}$ Dept. of Physics and Mathematics \& E.A.Milne Centre for Astrophysics, University of Hull, Cottingham Road, Kingston-upon-Hull, HU6 7RX\\
$^{33}$ Astronomy Unit, Queen Mary University London, Mile End Rd, London E1 4NS, UK \\ 
$^{34}$ School of Physics, The University of Melbourne, Parkville, VIC 3010, Australia \\
$^{35}$ SRON Netherlands Institute for Space Research, Landleven 12, 9747 AD, Groningen, The Netherlands}
\pubyear{2010} \volume{000}
\pagerange{\pageref{firstpage}--\pageref{lastpage}}

\maketitle
\label{firstpage}

\clearpage

\begin{abstract}
We present the GAMA Panchromatic Data Release (PDR) constituting over
230deg$^2$ of imaging with photometry in 21 bands extending from the
far-UV to the far-IR. These data complement our spectroscopic campaign
of over 300k galaxies, and are compiled from observations with a
variety of facilities including: GALEX, SDSS, VISTA, WISE, and
Herschel, with the GAMA regions currently being surveyed by VST and
scheduled for observations by ASKAP. These data are processed to a
common astrometric solution, from which photometry is derived for
$\sim 221,373$ galaxies with $r<19.8$ mag. Online tools are provided
to access and download data cutouts, or the full mosaics of the GAMA
regions in each band.

We focus, in particular, on the reduction and analysis of the VISTA
VIKING data, and compare to earlier datasets (i.e., 2MASS and UKIDSS)
before combining the data and examining its integrity. Having derived
the 21-band photometric catalogue we proceed to fit the data using the
energy balance code MAGPHYS. These measurements are then used to
obtain the first fully empirical measurement of the 0.1-500$\mu$m
energy output of the Universe. Exploring the Cosmic Spectral Energy
Distribution (CSED) across three time-intervals (0.3--1.1~Gyr,
1.1---1.8~Gyr and 1.8---2.4~Gyr), we find that the Universe is
currently generating $(1.5 \pm 0.3) \times 10^{35}$ h$_{70}$ W
Mpc$^{-3}$, down from $(2.5 \pm 0.2) \times 10^{35}$ h$_{70}$ W
Mpc$^{-3}$ 2.3~Gyr ago. More importantly, we identify significant and
smooth evolution in the integrated photon escape fraction at all
wavelengths, with the UV escape fraction increasing from 27(18)\% at
$z=0.18$ in NUV(FUV) to 34(23)\% at $z=0.06$.
The GAMA PDR can be found at: http://gama-psi.icrar.org/
\end{abstract}


\setlength{\extrarowheight}{0pt}

\section{Introduction}
Galaxies are complex systems. At the simplest level ionised gas cools
within a dark matter halo (White \& Rees 1978), condensing in the
densest environments to molecular hydrogen (Shu, Adams \& Lizano 1987)
which may become self-gravitating and lead to the formation of a
stellar population (Bate, Bonnell \& Bromm 2003). The stars replenish
the interstellar medium through supernovae, winds, and other mass-loss
processes (Tinsley 1980; Schoenberner 1983) leading to metal
enrichment, dust formation, and the heating of the interstellar medium
through shocks and other turbulent processes (McKee \& Ostriker 2007,
see also Fontanot et al.~2006). 

The dust attenuates (through absorption and scattering) a significant
portion of the starlight (Calzetti et al.~2000), up to 90\% depending
on inclination for disc systems (see Driver et al.~2007) and the
internal dust geometry and composition.  The absorbed fraction of the
UV/optical light (highly dependent on morphology but typically 30
percent for local Universe disk galaxies) is re-radiated at far-IR
wavelengths (Popescu \& Tuffs 2002; Tuffs et al.~2004; Driver et
al.~2008). Throughout this process gas is being drawn into the galaxy
from the intergalactic medium (Keres et al.~2005), outflows driven by
supernova expel material (Veilleux, Cecil \& Bland-Hawthorn 2005), and
tidal interactions with neighbouring dark matter halos may lead to
further mass-loss (Toomre \& Toomre 1972), or mergers (Lacey \& Cole
1993), as well as driving gas to the core leading to re-ignition of
the central super-massive black hole (Hopkins et al.~2006). In short,
galaxy evolution is governed by a very wide range of complex processes
that give rise to multiple energy production and recycling pathways
traced from X-ray to radio wavelengths.

Traditionally galaxy surveys have been predominantly single facility
campaigns (e.g., the SuperCOSMOS Sky Survey and other Digitised Plate
Surveys, Hambly et al.~2001; SDSS, York et al.~2000; 2MASS, Skrutskie
et al.~2006; IRAS, Soifer, Neugenbauer \& Houck.~1987; FIRST, White et
al.~1997; HIPASS, Barnes et al.~2001) and as a result only capable of
exploring a fairly narrow wavelength range. Therefore they often only
probe one constituent of this process, e.g., radio surveys which
sample the neutral gas content (Barnes et al.~2001), optical campaigns
sampling the stellar population (York et al.~2000), and far-IR
campaigns sampling the dust emission (Soifer et al.~1987). While
panchromatic datasets of relatively modest size have been constructed
(e.g., the Spitzer Infrared Nearby Galaxy Survey, Kennicutt et
al.~2003), they are generally too small to allow a full exposition of,
for example, environment and stellar-mass dependencies, or subdividing
samples to manage co-dependencies.

Part of the problem in assembling a {\it comprehensive} panchromatic
catalogue is the range of facilities required, which in many cases are
mismatched in sensitivities and resolutions. There are also
significant logistical issues: the physics underpinning the energy
processes at each wavelength are often very different; the distinct
data-streams often have very different wavelength dependent issues
requiring a broad range of specialist skills, and the lack of
cooperative global structures to coordinate observations across a
suite of facilities which cross international borders. Sampling the
full energy range therefore requires cooperation and collaboration
across a number of subject areas, the cooperation of time-allocation
committees, extensive resources to manage the many data-flows in an
optimal way, new techniques to combine the data in a robust manner,
and an open skies policy towards final data-products by national and
international observatories.

Progress in this area has mainly been driven by technological
advancements, coupled with large collaborative efforts, and
predominantly in two ways: (1) the construction of increasing samples
of well-selected nearby galaxies, often on an object-by-object basis
across the wavelength range (e.g., the Atlas of SEDs presented by
Brown et al.~2014 and the S$^4$G collaboration which now samples over
2000 galaxies, see Sheth et al.~2010 and Munoz-Mateos et al.~2015); or
(2) the concerted follow-up of the deep fields observed by the Hubble
Space Telescope (e.g., the HST GOODs, Giavalisco et al.~2004; HST
COSMOS, Scoville et al.~2007; and HST CANDLES, Grogin et al.~2011 and
Koekemoer et al.~2011, in particular). In the former the sample sizes
are modest ($\sim$100s---1000s of objects), in the latter the galaxies
sampled are predominantly at very early epochs (i.e., $z>1$). In short
no highly complete panchromatic catalogue of the nearby galaxy
population exists, suitable for comprehensive statistical analysis,
while also covering the full energy range.

The Galaxy And Mass Assembly survey (GAMA; Driver et al.~2009,~2011;
Baldry et al.~2010) is an attempt to provide a comprehensive
spectroscopic survey (Robotham et al.~2010; Hopkins et al.~2013; Liske
et al.~2015) combined with comprehensive panchromatic imaging from the
far-UV to far-IR and eventually radio. Results to date are based
mostly on the spectroscopic campaign combined with the optical imaging
to explore structure on kpc to Mpc scales, in particular the GAMA
group catalogue (Robotham et al.~2011), the filament catalogue
(Alpaslan et al.~2014), and structural studies of galaxy populations
(e.g., Kelvin et al.~2014).

Here we introduce the panchromatic imaging which has been acquired, by
us or other teams, over the past five years from a variety of ground
and space-based facilities. These surveys collectively provide
near-complete sampling of the UV to far-IR wavelength range, through
21 broad-band filters spanning from 0.15---500$\mu$m. The filters
represented are: FUV, NUV, $ugriz$, $ZYJHK_s$, W1, W2, W3, W4, 100$\mu
m$, 160$\mu m$, 250$\mu m$, 350$\mu m$, and 500$\mu m$. The
contributing surveys in order of increasing wavelength are: the GALEX
Medium Imaging Survey (Martin et al.~2005) plus a dedicated campaign
(led by RJT), the Sloan Digital Sky Survey Data Release 7 (Abazajian
et al.~2009), the VST Kilo-degree Survey (VST KiDS; de Jong et
al.~2013); the VIsta Kilo-degree INfrared Galaxy survey (VIKING; see
description of the ESO Public Surveys in Edge et al.~2013), the
Wide-field Infrared Survey Explorer (WISE; Wright et al.~2010), and
the Herschel Astrophysical Terahertz Large Area Survey
(Herschel-ATLAS; Eales et al.~2010). All of these facilities have
uniformly surveyed the four largest\footnote{GAMA's fifth region, G02,
  covers 20 sq deg and overlaps with one of the deep XXM XXL fields,
  see Liske et al.~(2015) for further details.}  GAMA regions referred
to as G09, G12, G15 and G23 (with only the latter field not covered by
SDSS). In the future the GAMA regions will be surveyed at radio
wavelengths by ASKAP (as part of the WALLABY or DINGO surveys) and at
X-ray wavelengths by eROSITA.

Combined, the four prime GAMA regions cover 230~deg$^{2}$ and have
uniform spectroscopic coverage to $r_{\rm Petro} < 19.8$ mag (G09,
G12, G15) or $i_{\rm Kron} < 19.2$ mag (G23), using a target catalogue
constructed from SDSS DR7 (G09, G12 and G15) or VST KiDS (G23)
imaging. The original GAMA concept is described in Driver et
al.~(2009), the tiling algorithm in Robotham et al.~(2010), the input
catalogue definition in Baldry et al.(2010), the optical/near-IR
imaging pipeline in Hill et al.~(2011), the spectroscopic pipeline in
Hopkins et al.~(2013), and the first two data releases including a
complete analysis of the spectroscopic campaign and redshift success,
in Driver et al.~(2011); and Liske et al.~(2015) respectively.

One of the scientific motivations is to assemble a comprehensive flux
limited sample of $\sim$221,000 galaxies with near-complete, robust,
fully-sampled spectroscopic coverage and robust panchromatic flux
measurements from the UV to the far-IR and thereafter apply spectral
energy distribution analysis codes to derive fundamental quantities
(e.g., stellar mass, dust mass, opacity, dust temperature,
star-formation rates etc).

In this paper we describe the processing and bulk analysis of the
panchromatic data and our discussion is divided into three key
sections. Section 2 outlines the genesis and unique pre-processing of
each imaging dataset into a common astrometric mosaic for each region
in each band (referred to hereafter as the GAMA SWarps), i.e.,
homogenisation of the data. Section 3 outlines our initial efforts
towards combining the various flux measurements from FUV to far-IR
which include a combination of aperture-(and seeing)-matched
photometry (SDSS/VIKING), table matching (GALEX, SDSS/VIKING, WISE),
curve-of-growth with automated edge detection (GALEX), and optical
motivated far-IR source detection (SDSS, SPIRE, PACS). In Section 4 we
demonstrate and test the robustness of the PDR.  Finally in Section 5
we provide an empirical measurement of the FUV-far-IR (0.1 ---
500$\mu$m) energy output of the Universe in three volume limited
slices centred at 0.5, 1.5, and 2.5~Gyr in look-back time. Note that
by energy output we refer to the energy being generated per Mpc$^{3}$
as opposed to the energy flowing through a Mpc$^3$ (e.g., Driver et
al.~2008,~2012; Hill et al.~2010). This is important as the former
refers to the instantaneous energy production rate of the Universe
(i.e., the luminosity density), whereas the latter is the integrated
energy production over all time, including the relic CMB photons
(e.g., Dom\'inquez et al.~2011).

Throughout this paper we use $H_0$=70$h_{70}$km s$^{-1}$ Mpc$^{-1}$ and
adopt $\Omega_M=0.27$ and $\Omega_{\Lambda}=0.73$ (Komatsu et
al.~2011). All magnitudes are reported in the $AB$ system.

\begin{figure*}
\centerline{\psfig{file=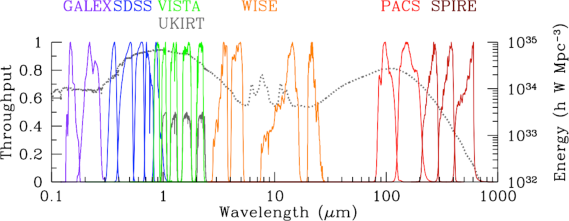,width=\textwidth}}

\caption{The 21 broad-band combined system throughput curves
  colour-coded by facility as indicated. Also shown (light grey line)
  is the recently measured(optical)/predicted(mid and far-IR) cosmic
  spectral energy distribution derived by Driver et al.~(2012). This
  CSED can be thought of as an energy weighted ``canonical'' galaxy
  spectral energy distribution and highlights how the GAMA PDR filter
  set samples the key energy regime for nearby and low redshift
  galaxies. Note filters are scaled to a peak throughput of 1 except
  UKIRT which are scaled to 0.5 for clarity.
\label{fig:filters}}
\end{figure*}

\begin{table*}
\caption{The GAMA panchromatic imaging regions \label{tab:fields}}
\begin{tabular}{ccccc} \hline \hline
GAMA region & SWarp RA centre & SWarp Dec centre & SWarp $\Delta$RA & SWarp $\Delta \delta$ \\ \hline
G09 & 09:00:30 & +00:15:00.0 & 19d15m24s & 7d30m18s \\
G12 & 11:59:30 & -00:15:00.0 & 19d15m24s & 7d30m18s \\
G15 & 14:29:30 & +00:15:00.0 & 19d15m24s & 7d30m18s \\ 
G23 & 23:00:00 & -32:30:00.0 & 14d00m00s & 6d00m00s \\ \hline
\end{tabular}

\noindent
Note: G02 is not included here but will be described in a dedicated release paper.
\end{table*}

\section{Panchromatic data genesis}
Fig.~\ref{fig:filters} shows the wavelength grasp of the 21
broad-band filters. The response curves represent the combined system
throughputs, normalised to a peak throughput of 1. Also shown as a
line (in light grey) is the nearby energy output from the combined
$z<0.1$ galaxy population derived from optical/near-IR analysis of the
GAMA dataset (see Driver et al.~2012). This highlights how the various
bands are sampling the stellar, polycyclic aromatic hydrocarbons
(PAHs), warm (temperature $\sim 50$K) and cool (temperature $\sim
20$K) dust emissions of the low redshift galaxy population (the curve
is shown for the energy output at $z=0$). In this section we start the
process of constructing individual spectral energy distributions
(SEDs) for every object within the GAMA main survey.

The first step is to place the diverse data onto a common astrometric
grid. Table.~\ref{tab:fields} defines the extent of the GAMA PDR
regions. We then use the Terapix SWarp package (see Bertin~2010) to
build single image mosaics for each waveband and each region (see Hill
et al.~2011).  The SWarp package uses the tangent plane (TAN) World
Coordinate System (WCS) to create a gnomic tangent-plane projection
centred on the coordinates shown in Table~1. One might argue about the
merit of constructing such large SWarped images ($\sim 110$~deg$^{2}$
each or up to 80GB for SDSS/VIKING data), however it was decided that
this was preferable to managing the $\sim 1$ million non-aligned
boundaries across the PDR. Taking each facility in turn we now
describe the pre-processing necessary to construct our GAMA
SWarps. Note that in addition to the native-resolution SWarps (see
Table~\ref{tab:data}) we also construct a set of SWarps at a common
3.39$''$ resolution (i.e., 10 times the VISTA pixel scale) for later
use in deriving coverage flags and background noise estimations.

\subsection{GALEX MIS, GO and Archive data}
The GALaxy Evolution eXplorer (GALEX, Martin et al.~2005) was a
medium-class explorer mission operated by NASA and launched on April
28$^{\rm th}$ 2003. The satellite conducted a number of major surveys
and observer motivated programs, most notably the all-sky imaging
survey (AIS; typically 200s integrations per tile) and the medium
imaging survey (MIS; typically 1500s per tile). The GALEX satellite is
built around a 0.5-m telescope with a field-of-view of 1.13~deg$^2$, a
pixel resolution of 1.5$''$, and a point-spread function FWHM of
4.2$''$ and 5.3$''$ in the FUV (153nm) and NUV (230nm) bands
respectively (Morrissey et al.~2007). Imaging data sampled at 1.5
arcsec from V7 of the GALEX pipeline forms the basis for constructing
the SWarped images. At the time of commencement of the GAMA survey the
GAMA regions contained patchy coverage with GALEX. A dedicated
programme led by one of us (RJT), was pursued providing further GALEX
observations to MIS depth (1500s) and completed in April 2013 (using
funds raised from the GAMA and Herschel-ATLAS Consortium to reactivate
and extend the GALEX mission). The final collated data provides
near-complete NUV and FUV coverage of the four primary GAMA
regions. Due to the failure of the FUV channel mid-mission, the
coverage at FUV in G23 is poor. However in G09, G12, and G15, coverage
is at the 90\% level in both bands (of which almost all is at MIS
depth in the NUV, and 60 percent is at MIS depth in the FUV; see
Section 2.6).

\begin{figure}

\centerline{\psfig{file=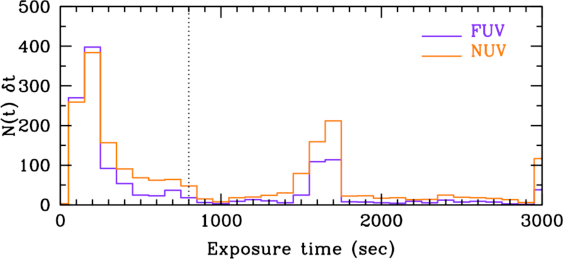,width=\columnwidth}}

\caption{The distribution of exposure times contributing to the final
  FUV and NUV SWarps. The dotted vertical denotes the cutoff below
  which frames are not used in the final SWarps. \label{fig:galex}}
\end{figure}

The analysis of the various GALEX datasets are described in detail in
Andrae (2014) and summarised in Liske et al.~(2015), and result in
background subtracted intensity maps scaled to the common GALEX
zero-points (Table~\ref{tab:sky}). As the data originate from a variety
of sources the exposure time is variable (see
Fig.~\ref{fig:galex}). To create our SWarps we take all available
GALEX data frames with exposure times greater than 800s. Within the
PDR only GALEX has such variable integration times.

~

In building the SWarps, a common circular mask (of radius 35$'$) was
used to trim the outer $\sim 5$ per cent of the image edges where the
data quality degrades due to the vignetting of the telescope aperture
(see Morrissey et al.~2007 and also Drinkwater et al.~2010 who adopted
a similar radius for the WiggleZ survey). In total we have 150, 137,
175, and 22 GALEX pointings in FUV and 167, 175, 176 and 133 in the
NUV for G09, G12, G15 and G23 respectively. These are combined to
produce single image SWarps at the FUV and NUV native resolution for
each region.

~

Note that a particular subtlety in building the FUV and NUV SWarps is
the nature of the sky backgrounds. In the FUV the majority of pixels
have zero flux (i.e., sky values of $<1$ photon) and hence the
distribution of sky pixel-values is highly asymmetrical (i.e.,
Poissonian). Great care was taken by the GAMA GALEX team (MS, RJT, EA)
to model and remove the backgrounds for each individual frame
appropriately and provide to GAMA background subtracted FUV data (see
Liske et al 2015, section 4.2 for further details). Hence when
constructing the FUV SWarps the background subtraction option was
switched off. Furthermore care should be taken in further background
analysis of GALEX FUV data by only using mean statistics and not
median statistics because of the highly asymmetrical background
distribution, and ensuring sufficient counts within any aperture to
derive a robust mean. The NUV data has a significant sky signal and is
therefore processed with the SWarp background subtraction on using a
$128 \times 128$ pixel mesh (i.e., 198$'' \times$198$''$).

\subsection{SDSS DR7}
The Sloan Digital Sky Survey (York et al.~2000) provides uniform
optical imaging of the G09, G12 and G15 regions in $ugriz$ bands at
0.4$''$ pixel resolution with a typical PSF FWHM of 1.4$''$ (see Hill
et al.~2011, figure 3). As the GAMA spectroscopic survey was
predicated on the SDSS imaging (Baldry et al.~2010) there is by design
uniform $ugriz$ coverage of the three equatorial GAMA regions (G09,
G12 and G15). In due course these regions, along with G23, are being
surveyed by the KiDS team which will provide both deeper (2mag)
and higher ($\times 2$) spatial resolution data (see de Jong et
al.~2013).

Here we re-utilise the large mosaic GAMA SWarps built from the Sloan
Digital Sky Survey Data Release 7 (Abazajian et al.~2009) by Hill et
al.~(2011, see update in Liske et al.~2015). In brief this involved
the construction of both native seeing SWarps and SWarps built from
data frames convolved to a uniform 2$''$ FWHM. The starting point is
to download all contributing SDSS frames from the DR7 database,
measure the PSF using PSFex (Bertin~2011), renormalise the data
to a common zero-point, and produce both native seeing and convolved
data frames (using {\sc fgauss} within {\sc HEASoft} to produce a
common PSF FWHM of 2$''$). We then build SWarps at both the native and
convolved resolutions from the distinct renormalised data
frames. During the SWarping process (see Bertin~2010; Hill et
al.~2011) the sky background is subtracted using a coarse
$512\times512$ pixel median filter to create a grid which in turn is
median filtered $3 \times 3$ before being fitted by a bi-cubic spline
to represent the background structure. The use of a large initial
median filter is to ensure minimal degradation of the photometry and
shapes of extended systems.

G23 lies too far south to be observed by SDSS but along with G09, G12
and G15 are being observed to a uniform depth within the KiDS survey.
The analysis of the KiDS data for GAMA and the preparation of the
input catalogue for G23 will be presented in Moffett et al.~(2015). At
the present time optical SWarps for G23 do not exist.

\subsection{VISTA VIKING}
The Visible and Infrared Telescope for Astronomy (VISTA, Sutherland et
al.~2015) is a 4.1m short focal length infrared optimised survey
telescope located 1.5km from the VLT telescopes at Paranal
Observatory.  VISTA is owned and operated by ESO and commenced
operations on 11$^{th}$ December 2009. VISTA then entered a five year
period of survey operation to conduct a number of ESO Public Surveys
(Arnaboldi et al.~ 2007). One of these surveys, the VIsta Kilo-degree
INfrared Galaxy Survey (VIKING), will cover 1500~deg$^2$ in two
contiguous regions located in the north and south Galactic caps plus
the G09 region. During the first two years of operations the VIKING
survey prioritised the GAMA and Herschel-ATLAS survey regions. The
VIKING survey footprint therefore covers all four primary GAMA regions
(by design), in five pass bands ($ZYJHK_s$) at sub-arsecond resolution
to projected $5 \sigma$ point-source sensitivities of 23.1, 22.3,
22.1, 21.5, 21.2 AB mag (respectively).

The near-IR camera (VIRCAM, Dalton et al.~2006) consists of 16
Raytheon VIRGO HgCdTe arrays (detectors) sampling an instantaneous
field-of-view of 0.6~deg$^{2}$ within the 1.65~deg diameter field. In
routine operation a set of micro-dithered and stacked frames are
formed, which are referred to as {\sc paw-prints}. The on-camera
dither sequence does not cover the gaps between the detectors and
hence a sequence of six interleaved {\sc paw-prints} is required to
produce a contiguous coverage rectangular {\sc tile} of 1.475~deg
$\times$ 1.017~deg.

{\sc Paw-print} data from the VISTA telescope is pipeline processed
(Lewis, Irwin \& Bunclark 2010) by the Cambridge Astronomy Survey Unit
(CASU) to produce astrometrically and photometrically calibrated
data. This process includes flat-fielding, bias subtraction, and
linearity corrections. The {\sc paw-prints} are then transmitted to
the Wide Field Astronomy Unit (WFAU) at the Royal Observatory
Edinburgh. The WFAU combines the {\sc paw-prints} into the {\sc tiles}
which are then served to the community through both the ESO archive
and the UK VISTA Science Archive (VSA). As the stacked {\sc tile} data
does not include sky-subtraction, sharp discontinuities can be
introduced into the {\sc tiles}. An additional concern is that the
{\sc tiles} may be constructed from {\sc paw-prints} taken during
significantly different seeing conditions. As we wish to both
sky-subtract and homogenise the point-spread function to allow for
aperture-matched photometry (see Hill et al.~2011), we requested
access to all the VIKING {\sc paw-print} data provided to the WFAU
from CASU, which lay within the GAMA primary regions. This consisted
of 9269 Rice compressed multi-extension {\sc fits} files (v1.3 data
from the CASU archive). These data were expanded out as individual
detectors resulting in 148304 individual frames. Properties were
extracted from the headers for each detector (airmass, extinction,
exposure time, zero point, sky level, seeing) and the seeing measured
directly using PSFex (Bertin~2011). The data for each
individual detector were then rescaled to a common zero point (30)
using Eqn.~1:
\begin{equation}
{\rm I}_{\rm New}= {\rm I}_{\rm Old} 10^{(-0.4(Z -2.5\log_{10}(1/t) 
- (\tau (\sec \chi - 1)) + X_{\rm V.AB}-30))}
\end{equation}
where $Z$ is the quoted zero-point, $t$ is the exposure time in
seconds, $\tau$ is the extinction in the relevant band and $\sec \chi$
is the airmass. These values are obtained directly from the {\sc fits}
headers post-CASU processing. $X_{\rm V.AB}$ is the conversion from
Vega to AB magnitudes (i.e., 0.521, 0.618, 0.937, 1.384 or 1.839 for
Z,Y,J,H,K respectively) and were derived by CASU from the convolution
of the complete system response functions convolved with the spectrum
of Vega and a flat AB spectrum. The response functions in comparison
to those for UKIRT are shown in Fig.~\ref{fig:filters}.

\begin{figure}

\centerline{\psfig{file=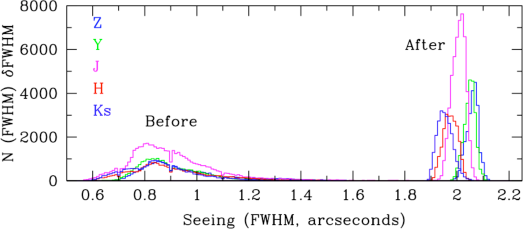,width=\columnwidth}}

\caption{Pre and post convolution seeing measurements of the 148304
  VISTA VIKING frames using PSFex. \label{fig:viking_seeing}}
\end{figure}

These data were convolved with the Gaussian kernel required to produce
a FWHM of $2''$ by assuming the PSF can be described as a Gaussian and
that the convolution of two Gaussians produces a broader Gaussian,
i.e., in line with our convolved SDSS data (see Hill et al.~2011).
Fig.~\ref{fig:viking_seeing} shows the pre- and post- convolved FWHM
as measured by PSFex. As can be seen the original seeing is
predominantly sub-arcsecond as expected from the ESO Paranal (NTT peak) site and
all the data lies well below our desired target PSF FWHM of
2$''$. Because the data is so much better than the target PSF FWHM
value the assumption of a Gaussian profile should produce
near-Gaussian final PSFs. Note that the $J$ band data is observed
twice, increasing the abundance of independent measurements. The
post-processed data is centred close to the target PSF FWHM of $2''$
with some indication of slight systematics between the bands at the
$5$ per cent level. Note this is not a major concern as we use
apertures with minimum major or minor diameters of $5''$ when
measuring our $u-K_s$ aperture-matched photometry (see Section
\ref{sec:utoK}).

\begin{figure}

\centerline{\psfig{file=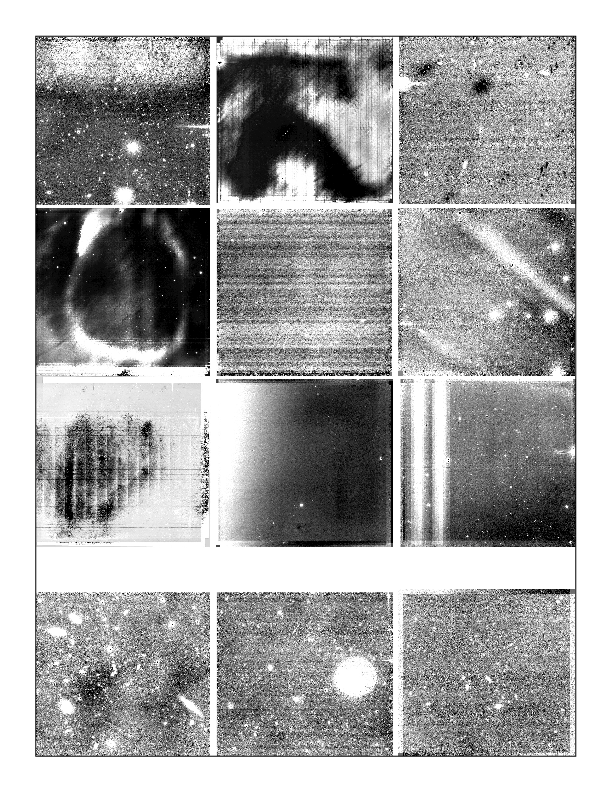,width=\columnwidth}}

\caption{Examples of poor quality (top three rows) and acceptable
  quality (bottom row) VIKING frames. Approximately 12 per cent of
  the VIKING data were visually inspected based on outlying values in
  airmass, sky background, zero-point, and
  seeing. \label{fig:viking_fails}}
\end{figure}

\begin{figure}

\centerline{\psfig{file=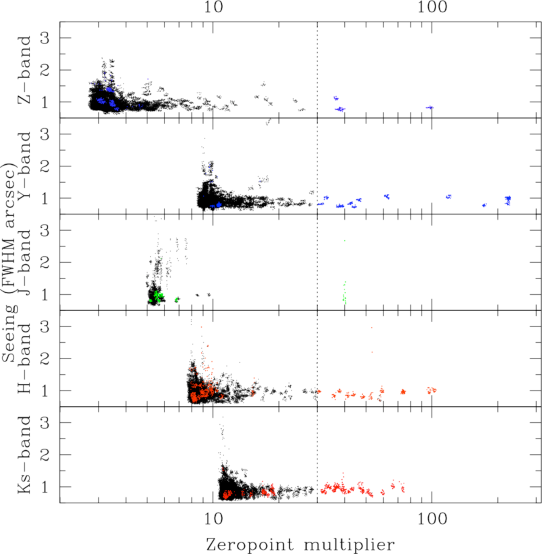,width=\columnwidth}}
\caption{Seeing versus zero-point multiplier for each band (as
  indicated). A cut of 30 appears to isolate the majority of low
  quality frames (indicated by the coloured
  points). \label{fig:viking_cuts}}
\end{figure}

From our initial SWarps we noted that a portion of data is clearly of
very low quality (see Fig.~\ref{fig:viking_fails}).  We therefore
elected to inspect a subset of the data by selecting three categories:
{\sc outliers} defined as those with seeing better than 0.5$''$ or
worse than 1.5$''$, a zero-point multiplier of greater than 40, a sky
value of less than 100 ADU counts or a CASU {\sc tilecode} not equal
to 0, 56, or -1, i.e., 9535 frames in total; {\sc control} defined as
a random set of 1000 frames not included in the above selection; and
{\sc sparse} defined as every detector 8 frame not already included in
one of the earlier samples, i.e., 6945 frames. These 16590 frames were
inspected by two of us (SPD, AHW) using the {\sc mogrify} routine
within the {\sc imagemagick} package to generate greyscale images
where the lowest 2 per cent of data were set black, the highest 10 per
cent white, and with histogram equalisation in-between. This scaling
amplifies background gradients rendering even the best quality data in
the poorest light (see Fig.~\ref{fig:viking_fails}). We then rejected
or accepted the frames via visual inspection and attempted to identify
a measurable quantity which best separated out the rejected frames,
see Fig.~\ref{fig:viking_cuts}. This resulted in the adoption of a
simple cut on the zero-point multiplier factor, whereby all frames
which require a rescaling of $\times 30$ or more are rejected in
addition to those already identified from the visual inspections. In
total 3262 of our 148304 frames were rejected (i.e., 2.2 per cent of
the data). Examples of accepted and rejected frames are shown in
Fig.~\ref{fig:viking_fails} and common causes are bright sky, detector
failures and telescope pointing errors.

\begin{figure}
\centerline{\psfig{file=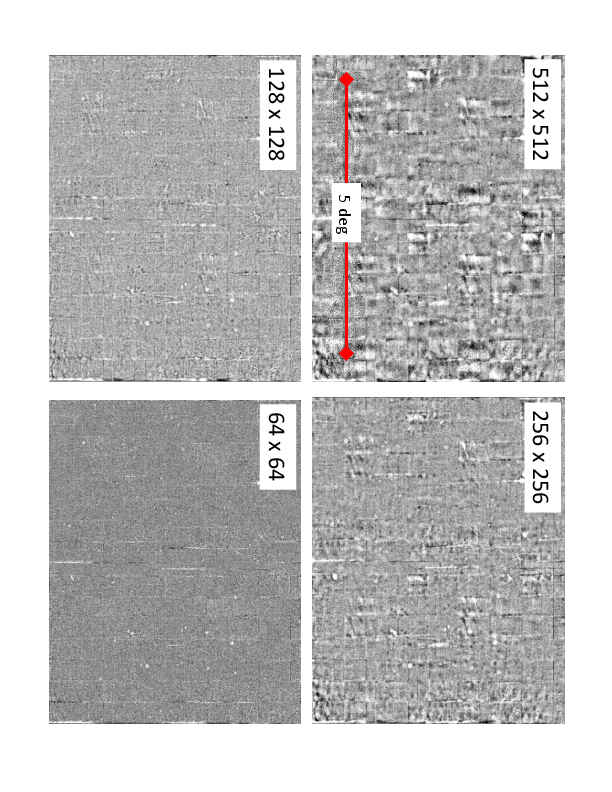,width=\columnwidth,angle=90}}
\caption{Examples of sections of VIKING data with various background
  subtractions as indicated.
\label{fig:viking_back}}
\end{figure}

\begin{figure}
\centerline{\psfig{file=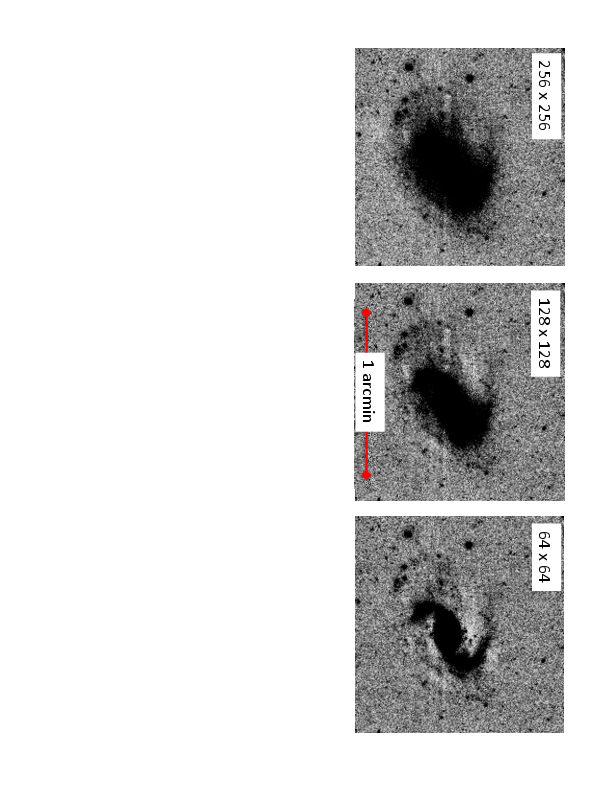,width=\columnwidth,angle=90}}

\vspace{-4.5cm}

\caption{An illustration of the impact of over-smoothing the
  background on extended objects. The galaxy shown is the largest
  system in the GAMA region, NGC0895 (located in G02). In the
  rightmost panel a significant portion of the galaxy has been removed
  due to the $64 \times 64$ pixel
  sky-subtraction process. \label{fig:viking_closeup}}
\end{figure}

\begin{figure}

\centerline{\psfig{file=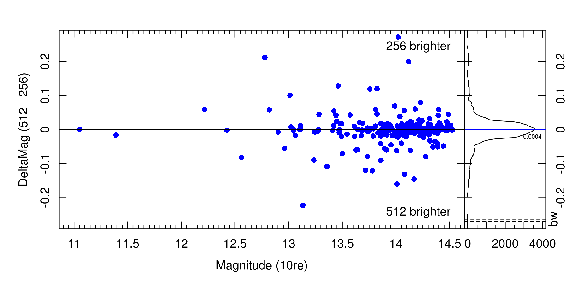,width=\columnwidth}}

\centerline{\psfig{file=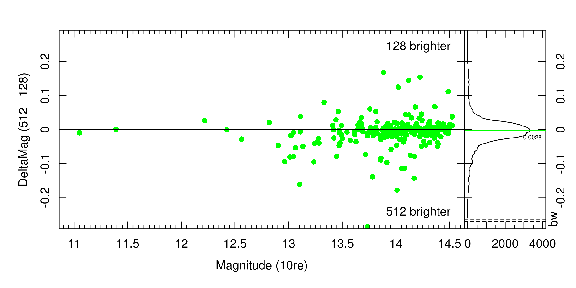,width=\columnwidth}}

\centerline{\psfig{file=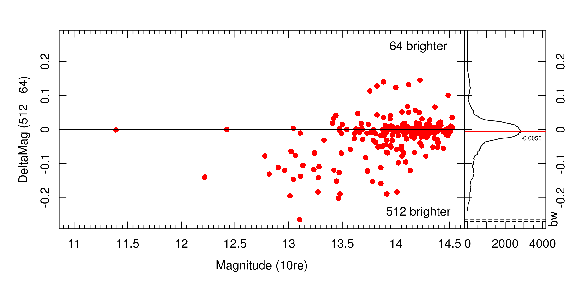,width=\columnwidth}}

\caption{A comparison of flux measurements of the brightest 100
  galaxies with varying background subtraction meshes. In each case
  the flux is compared against that derived from the $512 \times 512$
  pixel background mesh. In general it is only galaxies brighter than
  14$^{th}$ magnitude with the $64 \times 64$ background mesh whose
  photometry is compromised.  \label{fig:viking_mag}}
\end{figure}

\begin{figure}

\centerline{\psfig{file=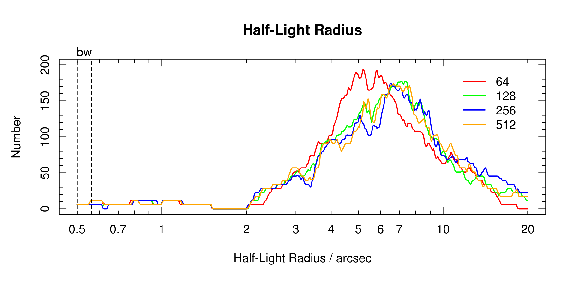,width=\columnwidth}}

\centerline{\psfig{file=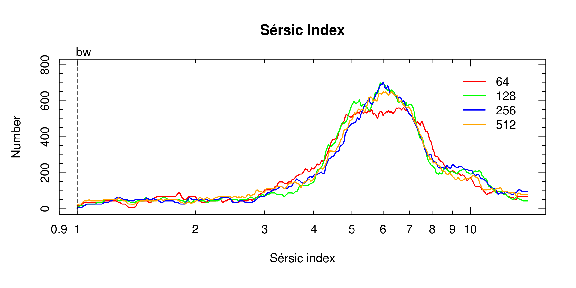,width=\columnwidth}}

\caption{The major-axis half-light radius (upper panel) and the
  S\'ersic index distributions (lower panel) for the brightest 100
  galaxies for various background mesh sizes as indicated.
\label{fig:viking_100}}
\end{figure}

The remaining frames were then SWarped (Bertin et al.~2010) to the
GAMA PDR regions specified in Table~\ref{tab:fields} with a pixel size
of 0.339$''$ using the TAN WCS projection. During the SWarping process
the background for each contributing detector was removed using a
$128\times 128$ pixel median filter which in turn was median filtered
by a $3 \times 3$ grid before being fitted with a bi-cubic spline. The
choice of background filter size is critical; too high and the
structure of the tiling becomes apparent in the SWarp (see
Fig.~\ref{fig:viking_back}), too low and galaxy photometry can be
affected (see Fig.~\ref{fig:viking_closeup}). To optimise the
background filter size we produced frames with a range of background
filter sizes and performed structural analysis of the brightest 100
galaxies using SIGMA (Kelvin et al.~2012).  Fig.~\ref{fig:viking_mag}
shows the magnitude offsets and Fig.~\ref{fig:viking_100} shows how
the measured major-axis half-light radii vary with background mesh
size. We tested pixel grids of $512\times512$, $256\times256$,
$128\times128$ and $64\times64$ and only the smallest filter size had
any noticeable impact on the measured properties and hence the second
smallest filter size was adopted. Note that this finer filtering
(compared to SDSS) is absolutely necessary because of (a) the mode of
observation (pointed v drift-scan) and (b) the higher-degree of sky
spatial variations in the near-IR wavebands.

\subsection{WISE}
The Wide-Field Infrared Survey Explorer (WISE; Wright et al.~2010) is
a medium-class explorer mission operated by NASA and was launched on
14$^{\rm th}$ December 2009. Following approximately 1 month of checks
WISE completed a shallow survey of the entire sky in 4 infrared bands
(3.4, 4.6, 12 and 22$\mu m$) over a ten month period. WISE is built
around a 40-cm telescope with a $47' \times 47'$ field-of-view, and
scans the sky with an effective exposure time of 11s per frame. Each
region of sky is typically scanned from tens to hundreds of times
(with fields further from the ecliptic being observed more
frequently). This allows the construction of deep stacked frames
reaching a minimum $5 \sigma$ point source sensitivity of 0.08, 0.11,
0.8 and 4~mJy in the W1(3.4$\mu m$), W2(4.6$\mu m$), W3(12$\mu m$) and
W4(22$\mu m$) bands (see Wright et al.~2010). The base ``Atlas'' data
consists of direct stacks and associated source catalogues which are
publicly available via the WISE and AllWISE data release hosted by the
Infrared Science Archive (IRSA). These public data have point-spread
function FWHM resolutions of $\sim 8.4'', 9.2'', 11.4''$ and $18.6''$
in W1, W2, W3 and W4 respectively and a 1.375$''$/pixel
scale. However, because of the stability of the point-spread function
of the WISE system, higher resolution can be attained using
deconvolution techniques, in particular ``drizzled'' co-addition and
the Maximum Correlation Method of Masci \& Fowler (2009; see Jarrett
et al.~2012). Here we use data which has been re-stacked via the
drizzle method as the MCM or HiRes method is computationally expensive
and only suited for very large nearby galaxies (see Jarrett et
al.~2012, 2013). In brief this involves:

~

\noindent
(1) gain-matching and rescaling the data ensuring a common photometric zero-point
calibration,

\noindent
(2) background level offset-matching,

\noindent
(3) flagging and outlier rejection,

\noindent
(4) co-addition using overlap area weighted interpolation and drizzle.

~

\begin{figure*}

\centerline{\psfig{file=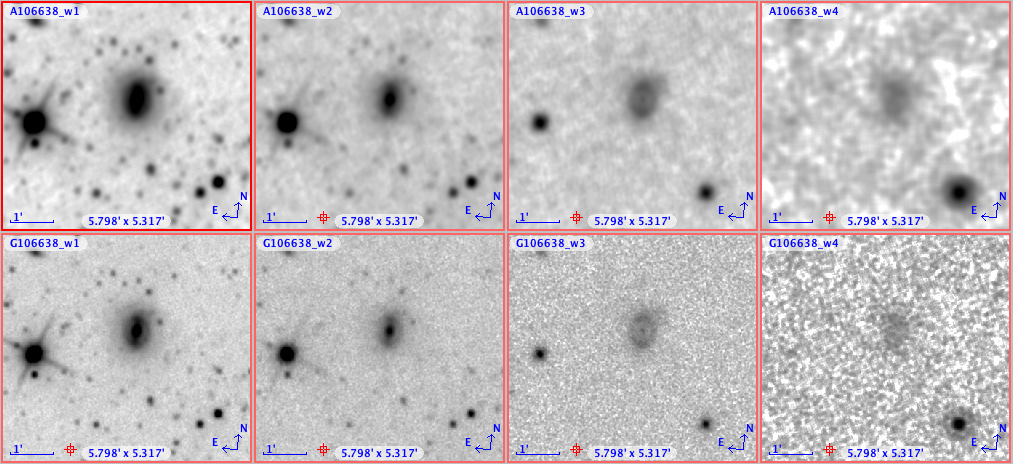,width=\textwidth,angle=0}}

\caption{A visual comparison of WISE Atlas images (upper rows) and
  WISE drizzled images (lower rows) for bands W1, W2, W3, W4
  (left-to-right). The panels are displayed over comparable ranges and
  the improvement in spatial resolution via the drizzling technique
  is self-evident. For details of the drizzling technique see Jarrett
  et al.~(2012). \label{fig:wisedrizzle}}
\end{figure*}

Here drizzle refers to the Variable Pixel Linear Reconstruction
technique of co-addition using a Point Response Function kernel to
construct the mosaics. Full details are provided in the WISE ICORE
documentation (Masci 2013). The drizzled data results in final
point-source FWHM of $5.9'', 6.5'', 7.0''$, and $12.4''$
(respectively), see Cluver et al.~(2014) and Jarrett et al.~(2012) for
further details. Fig.~\ref{fig:wisedrizzle} shows a comparison for one
of our GAMA galaxies between the ``Atlas'' and drizzled image in each
of the four bands. The ``drizzled'' frames are provided to the GAMA
team stacked, calibrated to a common zero point, and background
subtracted in sections of $1.56^{\circ} \times 1.56^{\circ}$. These
frames are then SWarped into a single large mosaic at the native pixel
resolution of $1''$ using the same field centre, and projection system
as for the previous datasets. In re-gridding the data we also include
the SWARP background subtraction using a $256 \times 256$ pixel
filter.

\subsection{Herschel-ATLAS}
The {\it Herschel} Space Observatory (Pilbratt et al.~2010) is
operated by the European Space Agency and was launched on May 14$^{\rm
  th}$ 2009 and conducted a number of major survey campaigns during
its 3.5 years of operation. The largest extragalactic survey, in terms
of areal coverage, is The Herschel Astrophysical Terahertz Large Area
Survey (Herschel-ATLAS; Eales et al.~2010). Herschel-ATLAS images were
obtained using Herschel's fast-scan parallel mode and covered
$\sim$600~deg$^2$ of sky in five distinct sky regions which included
the four principal GAMA fields.  The co-ordinated observations used
both the PACS (Poglitsch et al.~2010) and SPIRE (Griffin et al.~2010)
instruments to obtain scans at 100$\mu m$, 160$\mu m$, 250$\mu m$,
350$\mu m$, 500$\mu m$, i.e., sampling the warm and cold dust
components of galaxies from z=0 to z=4. The final maps were the
combination of two orthogonal cross-scans giving rise to PSFs with
Gaussianised FWHM of 9.6$''$ and 12.5$''$ in 100 and 160$\mu m$ and
18$''$, 25$''$ and 35$''$ in the 250, 350 and 500 $\mu m$ bands
respectively (see Valiante et al.~2015 for full details of the PSF
characterisation). The data were processed, calibrated, nebularized to
remove large scale fluctuations due to cirrus and large scale
clustering of high-z sources, (see Valiante et al.2015. and Maddox et
al.~2015), and finally mosaiced by the Herschel-ATLAS data reduction
team who provided the final maps and 5$\sigma$ source detection
catalogues. The reduction process for the two instruments are
described in detail in Ibar et al.~(2010), and Pascale et al.~(2011),
to be superseded shortly by Valiante et al.~(2015), and the method for
source detection is described in detail in Rigby et al.~(2011) also
updated in Valiante et al.~(2015). The absolute zero point calibration
is accurate to $\pm 10$ per cent for PACS and $\pm 7$ per cent for
SPIRE which provides a potential systematic pedestal in addition to
the random sky and object photon noise errors estimated later. Note
that as the Herschel-ATLAS data have not yet been publicly released
they remain subject to change. Every attempt will be made to ensure
the online GAMA PDR provides notifications of any changes or updates.

To date the Herschel-ATLAS data have been used to study the dust and
star-formation properties of both near and distant galaxies based on
far-IR/optical matched samples (see for example Dunne et al.~2011;
Smith et al.~2011, 2012; Bourne et al.~2012 Rowlands et al.~2012). To
pre-prepare the data for GAMA we re-SWarp the mosaics provided onto a
uniform grid using the field centres from Table~\ref{tab:fields} using
the TAN WCS projection, and preserving the original pixel size as
specified in the file headers and shown on Table~\ref{tab:data}.

\begin{table*}
\caption{Key meta-data information of the contributing
  datasets. \label{tab:data}}
\begin{tabular}{llcccccrrr} \hline \hline
Facility & Dataset & Instrument & Filter & Pivot    & Pixel      & Point-source & Frames   & $m_{\rm AB}-m_{\rm Vega}$ \\ 
         & or survey  & or technique & name    & Wavelength & Resolution & FWHM         & Supplied & (mag) \\ \hline
GALEX    & MIS+GO &  -    & FUV    & 1535\AA  & 1.5$''$ & 4.1$''$ & 279 & 2.16 \\
GALEX    & MIS+GO &  -    & NUV    & 2301\AA   & 1.5$''$ & 5.2$''$ & 297 & 1.67 \\ \hline
SDSS     & DR7  & - & $u$    & 3557\AA   & 0.339$''$ & 1.4$''$ & 26758 & 0.98 \\
SDSS     & DR7  & -      & $g$    & 4702\AA   & 0.339$''$ & 1.4$''$ & 26758 & -0.10 \\
SDSS     & DR7  & -      & $r$    & 6175\AA   & 0.339$''$ & 1.4$''$ & 26758 & 0.15 \\
SDSS     & DR7  & -      & $i$    & 7491\AA   & 0.339$''$ & 1.4$''$ & 26758 & 0.38 \\ 
SDSS     & DR7  & -      & $z$    & 8946\AA   & 0.339$''$ & 1.4$''$ & 26758 & 0.54 \\ \hline
VISTA    & VIKING & VIRCAM      & $Z$    & 8800\AA   & 0.339$''$ & 0.85$''$ & 15360 & 0.521 \\
VISTA    & VIKING & VIRCAM      & $Y$    & 10213\AA  & 0.339$''$ & 0.85$''$ & 15797 & 0.618 \\
VISTA    & VIKING & VIRCAM      & $J$    & 12525\AA  & 0.339$''$ & 0.85$''$ & 34076 & 0.937 \\
VISTA    & VIKING & VIRCAM      & $H$    & 16433\AA  & 0.339$''$ & 0.85$''$ & 15551 & 1.384 \\
VISTA    & VIKING & VIRCAM      & $Ks$    & 21503\AA  & 0.339$''$ & 0.85$''$ & 16340 & 1.839 \\ \hline
WISE    & AllSky & drizzled      & $W1$    & 3.37$\mu m$   & 1$''$ & 5.9$''$ & 40 & 2.683 \\
WISE    & AllSky & drizzled  & $W2$    & 4.62$\mu m$   & 1$''$ & 6.5$''$ & 40 & 3.319 \\
WISE    & AllSky & drizzled & $W3$    & 12.1$\mu m$   & 1$''$ & 7.0$''$ & 40 & 5.242 \\
WISE    & AllSky & drizzled & $W4$    & 22.8$\mu m$   & 1$''$ & 12.4$''$ & 40 & 7.871 \\ \hline
Herschel & ATLAS & PACS       & 100$\mu m$ & 101$\mu m$ & 3$''$ & 9.6$''$ & 4 (\& 1 for G23) & N/A \\
Herschel & ATLAS & PACS       & 160$\mu m$ & 161$\mu m$ & 4$''$ & 12.5$''$ & 4 (\& 1 for G23) & N/A \\
Herschel & ATLAS & SPIRE       & 250$\mu m$ & 249$\mu m$ & 6$''$ & 18$''$ & 4 (\& 1 for G23) & N/A \\
Herschel & ATLAS & SPIRE       & 350$\mu m$ & 357$\mu m$ & 8$''$ & 25$''$ & 4 (\& 1 for G23) & N/A \\
Herschel & ATLAS & SPIRE       & 500$\mu m$ & 504$\mu m$ & 12$''$ & 36$''$ & 4 (\& 1 for G23) & N/A \\ \hline
\end{tabular}
\end{table*}

\begin{table*}
\caption{Surface brightness limits of our GAMA SWarp set (FUV to mid-IR). \label{tab:sky}}
\begin{tabular}{cccccccr} \hline \hline
SWarp & Zero-Point & SWarp mean & $1 \sigma_{\rm Sky}$ & \multicolumn{2}{c}{5$\sigma$ limit} & Coverage \\ 
Facility/Filter/Field  & (AB mag for 1ADU) & (/$''$) & (mag arcsec$^{-2}$)$^{\dagger}$ & (mag)$^{\ddagger}$ & (Jy)$^\diamond$ & (\%)\\ \hline
GALEX   FUV   G09  &$  18.82  $&$  0.000148  $&$  28.41  $&$  25.23  $&$  2.94E-07  $& 88 \\
GALEX   FUV   G12  &$  18.82  $&$  5.04E-05  $&$  29.58  $&$  26.40  $&$  1E-07  $& 92 \\
GALEX   FUV   G15  &$  18.82  $&$  0.00013  $&$  28.54  $&$  25.37  $&$  2.59E-07  $& 95 \\
GALEX   FUV   G23  &$  18.82  $&$  0.000266  $&$  27.77  $&$  24.59  $&$  5.31E-07  $& 75 \\
GALEX   NUV   G09  &$  20.08  $&$  0.00125  $&$  27.35  $&$  23.92  $&$  9.84E-07  $& 94 \\
GALEX   NUV   G12  &$  20.08  $&$  0.00116  $&$  27.43  $&$  23.99  $&$  9.17E-07  $& 97 \\
GALEX   NUV   G15  &$  20.08  $&$  0.00161  $&$  27.07  $&$  23.64  $&$  1.27E-06  $& 95 \\
GALEX   NUV   G23  &$  20.08  $&$  0.00109  $&$  27.50  $&$  24.07  $&$  8.58E-07  $& 99 \\
 SDSS     $u$   G09  &$  30.00  $&$   142  $&$  24.61  $&$  22.24  $&$  4.61E-06  $& 100 \\
 SDSS     $u$   G12  &$  30.00  $&$   163  $&$  24.47  $&$  22.09  $&$  5.28E-06  $& 100 \\
 SDSS     $u$   G15  &$  30.00  $&$   156  $&$  24.51  $&$  22.14  $&$  5.06E-06  $& 100 \\
 SDSS     $g$   G09  &$  30.00  $&$  54.3  $&$  25.66  $&$  23.29  $&$  1.76E-06  $& 100 \\
 SDSS     $g$   G12  &$  30.00  $&$  65.3  $&$  25.46  $&$  23.09  $&$  2.12E-06  $& 100 \\
 SDSS     $g$   G15  &$  30.00  $&$  62.6  $&$  25.50  $&$  23.13  $&$  2.03E-06  $& 100 \\
 SDSS     $r$   G09  &$  30.00  $&$  76.7  $&$  25.28  $&$  22.91  $&$  2.49E-06  $& 100 \\
 SDSS     $r$   G12  &$  30.00  $&$  95.4  $&$  25.05  $&$  22.67  $&$  3.09E-06  $& 100 \\
 SDSS     $r$   G15  &$  30.00  $&$  90.5  $&$  25.10  $&$  22.73  $&$  2.93E-06  $& 100 \\
 SDSS     $i$   G09  &$  30.00  $&$   116  $&$  24.84  $&$  22.47  $&$  3.75E-06  $& 100 \\
 SDSS     $i$   G12  &$  30.00  $&$   140  $&$  24.62  $&$  22.25  $&$  4.56E-06  $& 100 \\
 SDSS     $i$   G15  &$  30.00  $&$   129  $&$  24.72  $&$  22.35  $&$  4.19E-06  $& 100 \\
 SDSS     $z$   G09  &$  30.00  $&$   506  $&$  23.23  $&$  20.86  $&$  1.65E-05  $& 100 \\
 SDSS     $z$   G12  &$  30.00  $&$   579  $&$  23.09  $&$  20.71  $&$  1.88E-05  $& 100 \\
 SDSS     $z$   G15  &$  30.00  $&$   556  $&$  23.13  $&$  20.76  $&$  1.81E-05  $& 100 \\
VIKING     Z   G09  &$  30.00  $&$  59.8  $&$  25.55  $&$  23.18  $&$  1.94E-06  $& 100 \\
VIKING     Z   G12  &$  30.00  $&$  60.6  $&$  25.54  $&$  23.17  $&$  1.97E-06  $& 100 \\
VIKING     Z   G15  &$  30.00  $&$    62  $&$  25.51  $&$  23.14  $&$  2.01E-06  $& 99 \\
VIKING     Z   G23  &$  30.00  $&$  67.9  $&$  25.41  $&$  23.04  $&$  2.2E-06  $& 100 \\
VIKING     Y   G09  &$  30.00  $&$   123  $&$  24.77  $&$  22.40  $&$  3.98E-06  $& 100 \\
VIKING     Y   G12  &$  30.00  $&$   110  $&$  24.89  $&$  22.52  $&$  3.56E-06  $& 100 \\
VIKING     Y   G15  &$  30.00  $&$   111  $&$  24.88  $&$  22.51  $&$  3.61E-06  $& 100 \\
VIKING     Y   G23  &$  30.00  $&$   129  $&$  24.71  $&$  22.34  $&$  4.2E-06  $& 100 \\
VIKING     J   G09  &$  30.00  $&$   167  $&$  24.44  $&$  22.06  $&$  5.43E-06  $& 100 \\
VIKING     J   G12  &$  30.00  $&$   161  $&$  24.48  $&$  22.10  $&$  5.23E-06  $& 100 \\
VIKING     J   G15  &$  30.00  $&$   146  $&$  24.58  $&$  22.21  $&$  4.74E-06  $& 100 \\
VIKING     J   G23  &$  30.00  $&$   155  $&$  24.52  $&$  22.14  $&$  5.04E-06  $& 100 \\
VIKING     H   G09  &$  30.00  $&$   329  $&$  23.70  $&$  21.33  $&$  1.07E-05  $& 98 \\
VIKING     H   G12  &$  30.00  $&$   302  $&$  23.79  $&$  21.42  $&$  9.82E-06  $& 99 \\
VIKING     H   G15  &$  30.00  $&$   313  $&$  23.75  $&$  21.38  $&$  1.02E-05  $& 97 \\
VIKING     H   G23  &$  30.00  $&$   325  $&$  23.71  $&$  21.34  $&$  1.06E-05  $& 100 \\
VIKING     K   G09  &$  30.00  $&$   332  $&$  23.69  $&$  21.32  $&$  1.08E-05  $& 100 \\
VIKING     K   G12  &$  30.00  $&$   337  $&$  23.67  $&$  21.30  $&$  1.09E-05  $& 100 \\
VIKING     K   G15  &$  30.00  $&$   303  $&$  23.79  $&$  21.42  $&$  9.83E-06  $& 100 \\
VIKING     K   G23  &$  30.00  $&$   285  $&$  23.86  $&$  21.48  $&$  9.25E-06  $& 100 \\
 WISE    W1   G09  &$  23.18  $&$  0.262  $&$  24.64  $&$  21.09  $&$  1.33E-05  $& 100 \\
 WISE    W1   G12  &$  23.18  $&$  0.281  $&$  24.56  $&$  21.01  $&$  1.43E-05  $& 100 \\
 WISE    W1   G15  &$  23.14  $&$  0.21  $&$  24.84  $&$  21.29  $&$  1.11E-05  $& 100 \\
 WISE    W1   G23  &$  23.14  $&$  0.187  $&$  24.96  $&$  21.41  $&$  9.9E-06  $& 100 \\
 WISE    W2   G09  &$  22.82  $&$  0.327  $&$  24.04  $&$  20.38  $&$  2.55E-05  $& 100 \\
 WISE    W2   G12  &$  22.82  $&$  0.367  $&$  23.91  $&$  20.26  $&$  2.87E-05  $& 100 \\
 WISE    W2   G15  &$  22.82  $&$  0.264  $&$  24.27  $&$  20.61  $&$  2.06E-05  $& 100 \\
 WISE    W2   G23  &$  22.82  $&$  0.229  $&$  24.42  $&$  20.77  $&$  1.79E-05  $& 100 \\ \hline
\end{tabular}

\noindent
$^{\dagger}$ $\mu_{1\sigma}=ZP-2.5\log_{10}({\rm \sigma_{\rm ADU}})$.

\noindent
$^{\ddagger}$ $5 \sigma~{\rm limit} = ZP-2.5\log_{10}(5 \sqrt{\pi
  HWHM^2} \sigma_{\rm ADU})$ where HWHM is Half Width Half-Maximum of
the seeing-disc (i.e., 0.5~FWHM).

\noindent
$^{\diamond}$ $F_{\nu}(Jy) = 3631 \times 10^{-0.4 {\rm mag}_{5 \sigma {\rm limit}}}$

\end{table*}

\begin{table*}
\caption{Surface brightness limits of our GAMA SWarp set (mid-IR to far-IR.}
\begin{tabular}{cccccccr} \hline \hline
SWarp & Zero-Point & SWarp mean & $1 \sigma_{\rm Sky}$ & \multicolumn{2}{c}{5$\sigma$ limit} & Coverage  \\ 
Facility/Filter/Field  & (AB mag for 1ADU) & (/$''$) & (mag arcsec$^{-2}$)$^{\dagger}$ & (mag)$^{\ddagger}$ & (Jy)$^\diamond$ & (\%) \\ \hline
 WISE    W3   G09  &$  23.24  $&$  2.33  $&$  22.32  $&$  18.59  $&$  0.000133  $ & 100 \\
 WISE    W3   G12  &$  23.24  $&$  2.68  $&$  22.17  $&$  18.44  $&$  0.000153  $ & 100 \\
 WISE    W3   G15  &$  23.24  $&$  1.77  $&$  22.62  $&$  18.89  $&$  0.000101  $ & 100 \\
 WISE    W3   G23  &$  23.24  $&$  2.18  $&$  22.39  $&$  18.66  $&$  0.000125  $ & 100 \\
 WISE    W4   G09  &$  19.60  $&$  0.278  $&$  20.99  $&$  16.64  $&$  0.000802  $ & 100 \\
 WISE    W4   G12  &$  19.60  $&$  0.305  $&$  20.89  $&$  16.54  $&$  0.000879  $ & 100 \\
 WISE    W4   G15  &$  19.60  $&$  0.208  $&$  21.31  $&$  16.96  $&$  0.000599  $ & 100 \\
 WISE    W4   G23  &$  19.60  $&$  0.265  $&$  21.05  $&$  16.69  $&$  0.000762  $ & 100 \\
 PACS   100   G09  &$  8.90  $&$  0.000562  $&$  -  $&$  12.96  $&$  0.0894  $ & 100 \\
 PACS   100   G12  &$  8.90  $&$  0.000545  $&$  -  $&$  12.99  $&$  0.0879  $ & 100 \\
 PACS   100   G15  &$  8.90  $&$  0.000547  $&$  -  $&$  12.99  $&$  0.0863  $ & 100 \\
 PACS   100   G23  &$  8.90  $&$  0.000476  $&$  -  $&$  13.14  $&$  0.0795  $ & 100 \\
 PACS   160   G09  &$  8.90  $&$  0.000278  $&$  -  $&$  13.44  $&$  0.103  $ & 100 \\
 PACS   160   G12  &$  8.90  $&$  0.000273  $&$  -  $&$  13.46  $&$  0.101  $ & 100 \\
 PACS   160   G15  &$  8.90  $&$  0.000271  $&$  -  $&$  13.47  $&$  0.101  $ & 100 \\
 PACS   160   G23  &$  8.90  $&$  0.000227  $&$  -  $&$  13.66  $&$  0.0903  $ & 100 \\
SPIRE$^{\flat}$   250   G09  &$  11.68  $&$  0.000759  $&$  -  $&$  12.56  $&$  0.0343  $ & 80 \\
SPIRE$^{\flat}$   250   G12  &$  11.68  $&$  0.00073  $&$  -  $&$  12.60  $&$  0.0330  $ & 81 \\
SPIRE$^{\flat}$   250   G15  &$  11.68  $&$  0.00073  $&$  -  $&$  12.59  $&$  0.0333  $ & 84 \\
SPIRE$^{\flat}$   250   G23  &$  11.68  $&$  0.000885  $&$  -  $&$  12.52  $&$  0.0357  $ & 100 \\
SPIRE$^{\flat}$   350   G09  &$  11.67  $&$  0.000447  $&$  -  $&$  12.36  $&$  0.0412  $ & 80 \\
SPIRE$^{\flat}$   350   G12  &$  11.67  $&$  0.000424  $&$  -  $&$  12.41  $&$  0.0394  $ & 81 \\
SPIRE$^{\flat}$   350   G15  &$  11.67  $&$  0.000423  $&$  -  $&$  12.41  $&$  0.0393  $ & 84 \\
SPIRE$^{\flat}$   350   G23  &$  11.67  $&$  0.000518  $&$  -  $&$  12.51  $&$  0.0357  $ & 100 \\
SPIRE$^{\flat}$   500   G09  &$  11.62  $&$  0.000228  $&$  -  $&$  12.16  $&$  0.0495  $ & 80 \\
SPIRE$^{\flat}$   500   G12  &$  11.62  $&$  0.000217  $&$  -  $&$  12.23  $&$  0.0467  $ & 81 \\
SPIRE$^{\flat}$   500   G15  &$  11.62  $&$  0.000221  $&$  -  $&$  12.21  $&$  0.0476  $ & 84 \\
SPIRE$^{\flat}$   500   G23  &$  11.62  $&$  0.000257  $&$  -  $&$  12.17  $&$  0.0490  $ & 100 \\ \hline
\end{tabular}

\noindent
$^{\dagger}$ $\mu_{1\sigma}=ZP-2.5\log_{10}({\rm \sigma_{\rm ADU}})$.

\noindent
$^{\ddagger}$ $5 \sigma~{\rm limit} = ZP-2.5\log_{10}(5 \sqrt{\pi HWHM^2} \sigma_{\rm ADU})$ where  HWHM is Half Width Half-Maximum of the seeing-disc (i.e., 0.5~FWHM).

\noindent
$^{\diamond}$ $F_{\nu}(Jy) = 3631 \times 10^{-0.4 {\rm mag}_{5 \sigma {\rm limit}}}$

\noindent
$^{\flat}$ SPIRE maps are in units of Jansky per Beam and to generate
these zero-points we have added a factor $2.5\log_{10}(B/N^2)$ where B is
the beam size given as 466, 821, and 1770 sq arcsec in 250, 350 and 500$\mu$m
respectively and N is the pixel size given in Table~\ref{tab:data}.
\end{table*}

\begin{figure*}
\centerline{\psfig{file=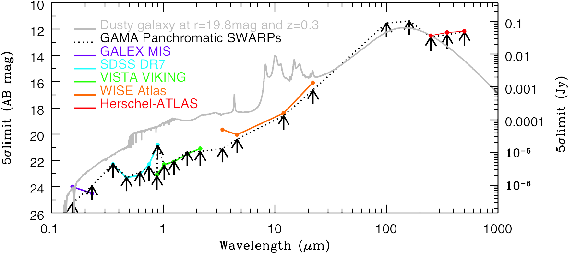,width=\textwidth}}

\caption{The sensitivity reached in each band as derived from the GAMA
  SWarps (black lines) and compared to the listed values (coloured
  lines). Also shown in grey is a typical SED for a dusty galaxy with
  $r_{\rm AB}=19.8$ mag. \label{fig:depths}}
\end{figure*}

\subsection{Cosmetic and noise characteristics of the GAMA SWarp set}
To assess the quality of GAMA SWarps we derive the background noise
distributions (i.e., sky-subtracted), within selected regions for each
of our LOW-RES (i.e., 3.39$''$) SWarps, which are displayed from
$-2\sigma$ to $+2\sigma$ in Figs.~A\ref{fig:g09} to A\ref{fig:g23} for
G09, G12, G15 and G23 respectively. The black rectangle represents the
GAMA region and the dotted blue rectangle the selected region from
which the noise characteristics are derived (the mode and
3$\sigma$-clipped standard deviation). These images show no obvious
major sky gradients across the sky regions, however, they do show
interesting substructure which highlights correlations in the
underlying noise properties. In most cases the correlations highlight
the genesis, i.e., the SDSS stripes, GALEX pointings, and VIKING
paw-prints. In these cases the noise properties for each particular
frame/scan is dictated by the conditions during observations (SDSS and
VISTA) or the variability of the various integration times
(GALEX). While uniform backgrounds are highly desirable, these are
never achieved in practice.  Some SDSS scans will be slightly less
noisy than others and some paw-prints will have significantly
amplified noise characteristics. Interestingly the WISE and
Herschel-ATLAS data show the least structure which mainly reflects the
benefits of using fixed integration times as well as operating outside
the confines of a time-varying atmosphere. However, some impact of
observing close to the moon is apparent in the WISE G12 SWarps. Also
noticeable in the Herschel-ATLAS data is the reduced noise in the
overlap regions as expected.

The noise distributions derived from the GAMA SWarps are shown in
Table~\ref{tab:sky}, for GALEX, SDSS, VISTA and WISE data these
statistics are derived from fitting a Gaussian distribution to the
histogram of data values below the mode. They therefore do not include
any confusion estimate and assume the noise is uncorrelated. In all
cases the distributions are very well described by a normal
distribution implying that the systematic frame-pistoning (i.e., ZP
offsets) in the data (arising from the independent calibration of the
distinct pointings), is operating at a relatively low level and within
the range of the pixel-to-pixel variations. Using the
$3\sigma$-clipped standard deviations we derive (analytically) the
1$\sigma$ surface brightness limits and the 5$\sigma$ point-source
detection limits for each of the SWarp images (see
Table.~\ref{tab:sky}).  For the PACS and SPIRE data, where correlated
noise is believed to be an issue, we derive the 5$\sigma$ detection
limits directly by placing apertures equivalent to the Beam size at
random locations across the SWarps and measuring the standard deviation
of the resulting aperture fluxes (again fitting to the distribution
below the mode). In Fig.~\ref{fig:depths} the GAMA SWarp detection
limits are compared to the values listed online for each facility (as
indicated by the colour lines). For GALEX MIS, SDSS DR7, and VIKING,
the depths probed agree extremely well. Note that our derived WISE W1
band limit appears significantly deeper than that quoted by the WISE
collaboration this is because our values ignore confusion (i.e., fits
to the negative noise distribution) whereas the WISE quoted value
incorporates this aspect. For Herschel-ATLAS SPIRE we note that the
agreement with the SPIRE values reported in Valiante et al.~(2015) is
extremely good.

\begin{figure*}
\centerline{\psfig{file=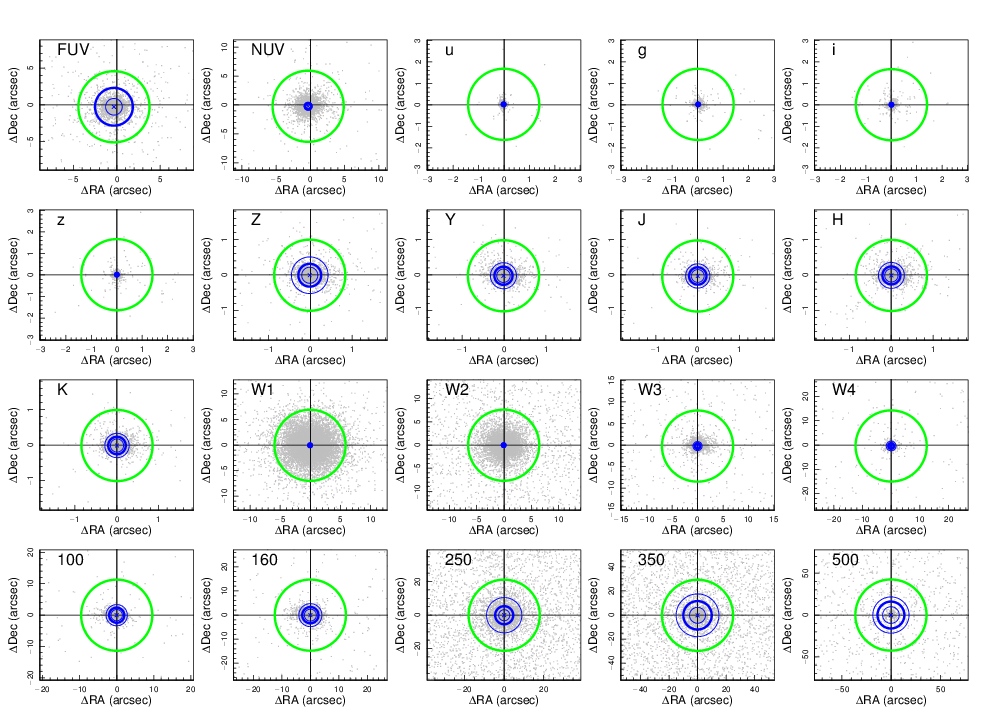,width=\textwidth}}

\caption{Confirmation of the astrometric accuracy. Each diagram shows
  the positional offsets of that particular band against the $r$ band
  GAMA Input or Tiling catalogues (grey data points). The centroid of
  the population is shown with a blue cross and the (native) PSF FWHM
  is shown as a green circle. The circles enclosing 50, 66 and 80 per
  cent of the population is shown by a thin, thick and thin blue line
  respectively. In all cases the relative astrometry is robust to
  $<0.1$PSF FWHM, and the 66 per cent spread enclosed with $0.5\times$
  the PSF FWHM. \label{fig:astrom}}
\end{figure*}

\subsection{Astrometric verification}
To check the astrometric alignment of the SWarps we run SExtractor
over either the entire SWarp set (GALEX, WISE, and Herschel) or a 4sq
deg section from G12 (SDSS and VIKING, to keep CPU requirement
manageable). We use relatively high signal-to-noise cuts of 100
(GALEX, where the data is not uniform), 10 (WISE and Herschel), or 3
(SDSS and VIKING). We then match to either the GAMA InputCat with $r <
17.0$ mag (GALEX, SDSS, VIKING and WISE) or the GAMA TilingCatv45 with
$r < 17.0$ mag and $(u-g) < 1.5$ (Herschel-ATLAS) to isolate
star-forming galaxies. Fig.~\ref{fig:astrom} shows the resulting
$\Delta$ RA and $\Delta$ Dec diagrams for each band compared to the
canonical $r$-band data (grey data points). On Fig.~\ref{fig:astrom}
the blue cross (mostly not visible) defines the centroid and the thick
green circle indicates the PSF FWHM for that band. The thick blue band
defines the region which encloses 66 percent of the population (after
accounting for the density of random mis-matches), and the thin blue
circles enclose either 50 percent or 80 percent of valid
matches. Fig.~\ref{fig:astrom} highlights that in all cases the
centroid of the RA and Dec offset is extremely close to zero (below
0.3$''$ in all bands with the FUV and NUV showing the largest
offsets, and below 0.02$''$ in the optical and near-IR), and that the
66 percent sprawl lies within $0.5\times$ the PSF FWHM in all
bands. We therefore consider the astrometry to be as one would expect
given the respective FWHM seeing values.

\begin{figure}

\psfig{file=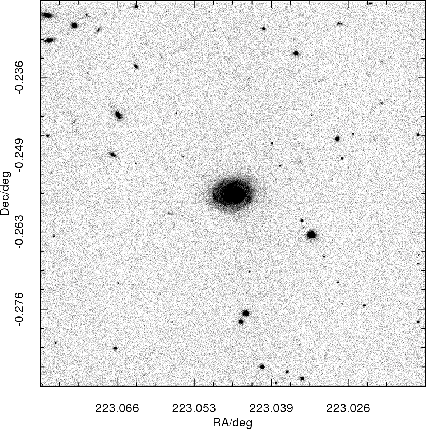,width=4.0cm}
\psfig{file=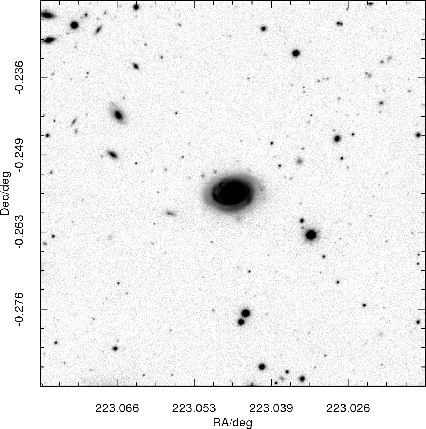,width=4.0cm}

\caption{A comparison of the quality of the SDSS $z$ band data (left) against the VISTA VIKING $Z$ band (right). \label{fig:zbandcomparison}}

\end{figure}

\begin{figure}

\psfig{file=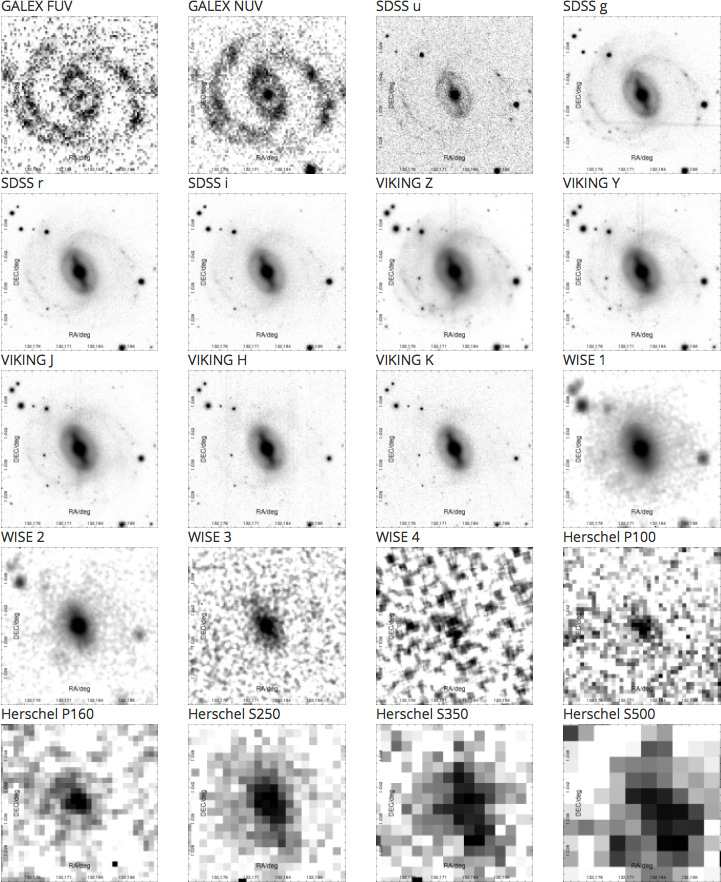,width=\columnwidth}

\caption{20 band panchromatic imaging for a $1' \times 1'$ region
  centred on GAMA galaxy G371633. Filters increase in wavelength
  proceeding from left to right and top to bottom, note the SDSS $z$
  filter is omitted, i.e., FUV, NUV, u, g, r, i, Z, Y, J, H, K, W1, W2,
  W3, W4, 100, 160, 250, 350, 500. 
Produced using the GAMA $\Psi$: http://gama-psi.icrar.org/
\label{fig:examples}}

\end{figure}

\begin{figure}

\psfig{file=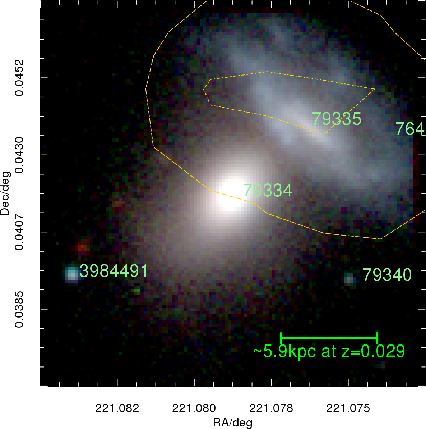,width=4.0cm}
\psfig{file=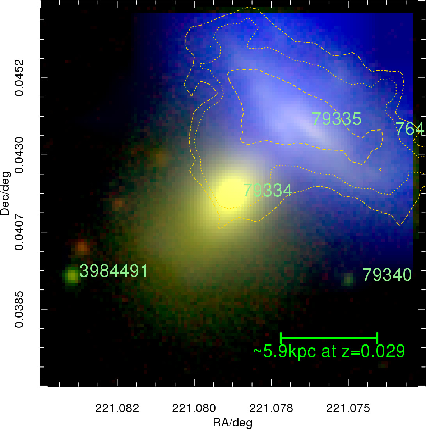,width=4.0cm}

\psfig{file=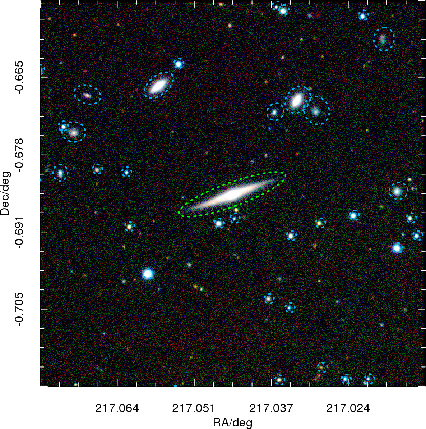,width=4.0cm}
\psfig{file=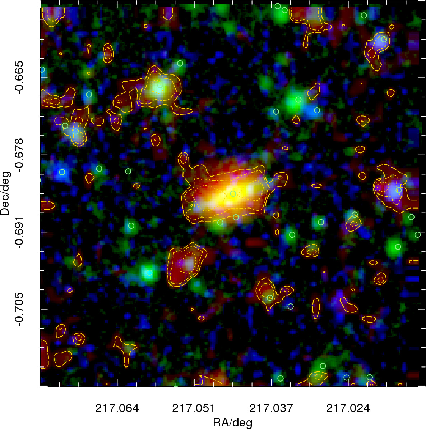,width=4.0cm}

\caption{Upper left: A colour composite image of G79334 produced by
  combining the SDSS $g$ \& $r$ with the VIKING H band
  images. Overlaid are the contours from SPIRE-250 band. GAMA IDs are
  marked. Upper right: A colour composite of G79334 from GALEX NUV,
  VIKING $Z$ \& $K_s$ and with contours overlaid from WISE.  Lower
  left: A $2' \times 2'$ colour composite centred on G48432 made from
  data extracted from the GAMA SDSS $r$ \& $i$ SWarp combined with the
  VIKING $H$ SWarp and with the apertures used for the
  aperture-matched photometry overlaid. (lower right) A composite colour image
  of G48432 made from GALEX FUV (blue channel), WISE W1 (green channel) and SPIRE
  250$\mu m$ (red channel and contours).  All images produced using the
  online GAMA $\Psi$ tool: http://gama-psi.icrar.org/
\label{fig:examples2}}

\end{figure}

\subsection{Visual inspection of the combined data and data access}
Our full dataset is diverse and the volume large. In order to inspect the data
we have developed a publicly available online tool which provides both
download links to the individual SWarps, as well as an option to
extract image regions from the dataset. Users can also build RGB
colour images using any of the 21 bands as well as overlay contours
and basic catalogue information (e.g., GAMA IDs, photometry apertures,
and object locations). The GAMA Panchromatic SWarp Imager ($\Psi$) is
therefore extremely versatile and useful for exploring the data
volume:

\noindent
\url{http://gama-psi.icrar.org/}

~

\noindent Fig.~\ref{fig:zbandcomparison},
~\ref{fig:examples}~\&~\ref{fig:examples2} show examples of various
extractions using the tool with Fig.~\ref{fig:zbandcomparison} showing
the significant increase in depth from the SDSS $z$ band data to the
VISTA VIKING $Z$ band. Fig.~\ref{fig:examples} shows a single GAMA
galaxy in 20 of the 21 bands (note that the SDSS $z$ band is not shown
here), and Fig.~\ref{fig:examples2} shows various colour combinations
with contours, IDs and apertures overlaid as indicated.  Note that
searches can be made based on GAMA ID {\it or} RA and Dec and is
therefore of use to high-z teams with objects in the GAMA regions
(e.g., Herschel-ATLAS team).

~

Using GAMA $\Psi$ via the link above, one can also access the
individual SWarps files including the native, convolved and 
weight-maps and the XML files which contain, the pixel data, a description of
the weights, and a listing of the constituent files making up the
SWarp respectively. The weight-maps are particularly useful and can be used to
determine both the coverage and provide a mask. Zero values in the
weight SWarp imply no data while non-zero values imply coverage. These
weight-maps have been used to generate the coverage statistics shown
in col.~7 of Table.~\ref{tab:sky}. The SWarps, weight-maps, and XML
files can all be downloaded from:

\noindent
\url{http://gama-psi.icrar.org/panchromaticDR.php}

~

\noindent
However, note that files sizes vary from 100KB (for XML files) to
up to 80GB (for SWarps and weight-maps).

\section{Panchromatic photometry for G09, G12 and G15 only}
Vital to successful analysis of panchromatic data are robust flux
measurements, robust errors, and a common deblending solution. This is
particularly difficult when the flux sensitivities and spatial
resolutions vary significantly, as is the case with the GAMA PDR (see
Figs~\ref{fig:depths},~\ref{fig:examples}~\&~\ref{fig:examples2},
i.e., 35$''$ to 0.7$''$ spatial resolution). In an ideal situation one
would define an aperture in a single band and then place the same
aperture at the same astrometric location in data with identical
spatial sampling. This is the strategy we pursued in Hill et
al.~(2011, see also Driver et al.~2011) to derive $u-K_s$
aperture-matched photometry (using the seeing-convolved SWarps
convolved to a 2$''$ FWHM). While we can still implement this strategy
in the $u$ to $K_s$ range (see Section 4.2 below) we cannot easily
extend it outside this wavelength range because of the {\it severe}
resolution mismatch (see Table~\ref{tab:data}). Software ({\sc
  LAMBDAR}) is being developed to specifically address this issues and
will be described in a companion paper (Wright et al.~2015). In the
meantime, we assemble a benchmark panchromatic catalogue from a
combination of aperture-matched photometry, table matching, and
optically-motivated (forced) photometry. It is worth noting that the
GAMA PDR assembled here while heterogeneous across facilities is
essentially optimised for each facility, and therefore optimal for
studies not requiring broad panchromatic coverage.

In the FUV and NUV, we use the GAMA GALEX catalogue described in Liske
et al.~(2015) and which uses a variety of photometry measures
including curve-of-growth and the GALEX pipeline fluxes. In the
optical and near-IR we apply the aperture-matched method mentioned
above and described in detail in the next sections. In the mid-IR we
use the WISE catalogues described in Cluver et al.~(2014). Longwards
of the WISE bands we adopt a strategy developed by the Herschel-ATLAS
team (Bourne et al.~2012, see Appendix A) to produce
optically-motivated aperture measurements (sometimes referred to as
forced photometry). This is applied to all GAMA targets which lie
within the PACS and SPIRE 100 to 500$\mu$m data.

\subsection{Aperture-matched photometry from $u$ to $K_s$: IOTA \label{sec:utoK}}
The $u$ to $K_s$ band data has been convolved to a common $2''$ FWHM
seeing (see Fig.~\ref{fig:viking_seeing}). For each object in the GAMA
tiling catalogue with a secure redshift (TilingCatv44, i.e., a valid
galaxy target within the specified regions with $r_{AB} < 19.8$ mag,
see Baldry et al.~2010) we perform the following tasks:

~

\noindent
(1) extract a $1001 \times 1001$ pixel region in all 10 bands ($ugrizZYJHK_s$),

\noindent
(2) run SExtractor in dual object mode with $r$ as the primary band,

\noindent
(3) identify the SExtractor object closest to the central pixel ($2''$ max),

\noindent
(4) extract the photometry for this object in the two bands,

\noindent
(5) repeat for all bands.

~

In essence this process relies on SDSS DR7 for the initial source
detection and initial classification including an $r$-band Petrosian
flux limit to define the input catalogue. However, the final
deblending and photometry is ultimately based on SExtractor (using the
parameters described in Liske et al.~2015 optimised for our convolved
data). An identical aperture and mask and deblend solution ---
initially defined in the $r$ band --- is then applied to the
$ugizZYJHK_s$ bands. In order to manage this process efficiently for
220k objects we use an in-house software wrapper, IOTA.

\subsection{Recalibration of the $u$ to $K_s$ photometry}
\begin{figure*}

\centerline{\psfig{file=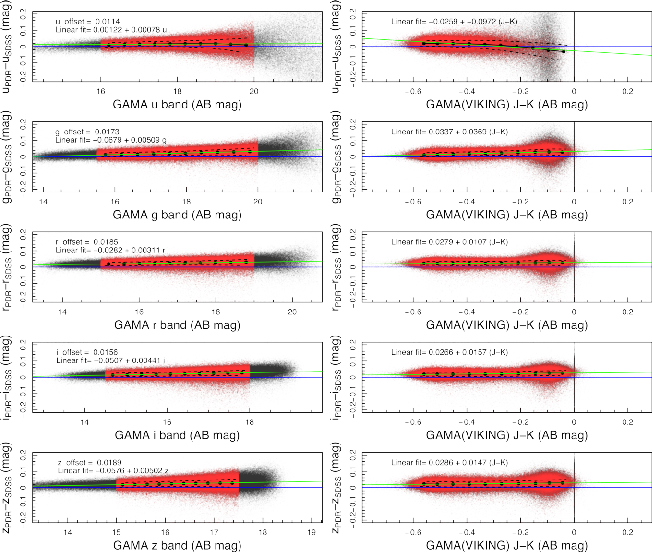,width=\textwidth}}

\caption{Comparison between SDSS PSF photometry versus SDSS (IOTA) for
  420k stars.  Highlighted in red are data deemed to lie in the flux
  and colour regions for which comparisons can be made.
\label{fig:stars_opt}}
\end{figure*}

\begin{figure*}

\centerline{\psfig{file=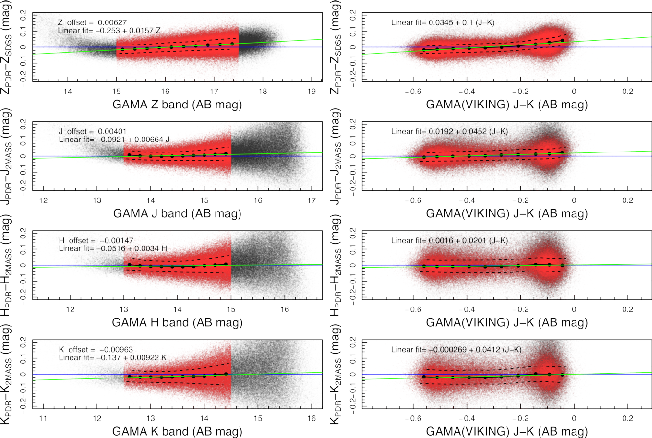,width=\textwidth}}

\caption{Comparison between 2MASS stellar photometry versus VIKING (IOTA)
  for 420k stars.  Highlighted in red are data deemed to lie
  in the flux and colour regions for which comparisons can be made.
\label{fig:stars_nir}}
\end{figure*}

The VIKING data is relatively new and to assess the absolute zero-point
errors we test the consistency of the photometry between our measured
VIKING data and the 2MASS point source catalogue. We achieve this by
extracting all catalogued stars in the extended GAMA regions from
InputCatv06 which itself is derived from SDSS DR7 (see Baldry et
al.~2010). To obtain near-IR flux measurements we uploaded the objects
classified as stars (see Baldry et al.~2010) to the IPAC Infrared
Science Archive (IRSA) and queried the 2MASS All-Sky Point Source
Catalogue (on 2013-06-07). We obtained 498,637 matches for which
photometry existed in 1 or more of the 2MASS bands ($JHK_s$). This
sample was trimmed to the exact GAMA RA extents to produce
catalogues of 201671, 92224 and 131976 stars in G09, G12, and G15
respectively. We ran {\sc IOTA} on these objects to derive
$ugrizZYJHK_s$ photometry based on Kron apertures with a minimum
aperture diameter of $5''$. Fig.~\ref{fig:stars_opt} \&
\ref{fig:stars_nir} shows the resulting zero-point comparisons versus
magnitude (left panels) and versus the VIKING $(J-K)_{AB}$ colour
(right panels) for filters $ugrizZJHK_s$ (top to bottom)
respectively. Note that for the $ugriz$ bands we compare directly to
SDSS PSF mags corrected to AB (i.e., $u_{\rm AB}=u_{\rm SDSS}-0.04$
and $z_{\rm AB}=z_{\rm SDSS}+0.02$) for the $ZJHK_s$ bands we convert
the 2MASS mags into the VISTA passband system, using the colour
transformations derived by the VISTA Variables in the Via Lactea
Survey (VVV) team (Soto et al.~2013) which are:
\begin{eqnarray}
J_{\rm VISTA}=J_{\rm 2MASS}-0.077(J_{\rm 2MASS}-H_{\rm 2MASS}) \\
H_{\rm VISTA}=H_{\rm 2MASS}+0.032(J_{\rm 2MASS}-H_{\rm 2MASS}) \\
K_{\rm VISTA}=K_{\rm 2MASS}+0.010(J_{\rm 2MASS}-K_{\rm 2MASS}) 
\end{eqnarray}
Finally we implement the Vega to AB correction appropriate for the
VISTA filters, see Table.~\ref{tab:data}.

At brighter magnitudes the deeper VIKING data will suffer from
saturation, and at fainter magnitudes the shallow 2MASS data will
become swamped by noise. Fig.~\ref{fig:stars_opt} and
\ref{fig:stars_nir} show the direct comparisons (black data points),
and the data which we consider robust to saturation and limiting
signal-to-noise (red data points). Also shown on the figures are the
derived global offset values (blue lines), and the simple linear fits
(green lines) to the medians (black squares with errorbars) for both
the magnitude (left) and colour comparisons (right). The dotted lines
indicate the quartiles of the data.

We conclude that the absolute zero-point calibration is robust across
the board to $\pm 0.02$ mag within the magnitude and colour ranges
indicated (red data points). However, it is extremely important to
recognise that the majority of our galaxies lie significantly outside
the flux and colour ranges which we are examining here. As we shall
discuss in Section~\ref{sec:linearity} this can cause significant and
intractable issues. To quantify the potential for zero-point drift
between the calibration regime and operating regime we show in
Table~\ref{tab:zperror} possible zero-point offsets one might derive
at the typical flux and the typical colour of the GAMA sample using
either (a) a simple offset (Fig.~\ref{fig:stars_opt} \&
\ref{fig:stars_nir} blue line), (b) a linear fit with magnitude (the
linear fit shown as a green line in Fig.~\ref{fig:stars_opt} \&
\ref{fig:stars_nir} left panels) and, (c) a linear fit with colour
(the linear fit shown as a green line on Fig.~\ref{fig:stars_opt} \&
\ref{fig:stars_nir} right panels). Any one of these relations, or some
combination of, could be valid and hence the range reflects the
uncertainty in the absolute zero-point calculations for our
filters. We elect not to correct our data using any of these
zero-points but instead incorporate the possible systematic zero-point
error (indicated in the final column) into our analysis.

\begin{table*}
\caption{Zero-Point uncertainties in each band. \label{tab:zperror}}
\begin{tabular}{c|c|c|c|c|c|c|c} \hline \hline
Band & GAMA Median  & GAMA Median & \multicolumn{3}{c|}{Potential zero-point (ZP) offsets} & ZP unc. & Adopted \\
     & flux limit (mag) & ($J-K_s$) colour (mag) & Absolute & linear with mag & linear with colour & Mean $\pm$ Std. & ZP error \\ \hline
u & 21.48 & 0.37 & +0.011 & +0.018 & -0.020 & $+0.003 \pm 0.020$ & 0.02 \\
g & 20.30 & 0.40 & +0.017 & +0.035 & +0.048 & $+0.033 \pm 0.016$ & 0.05 \\
r & 19.35 & 0.41 & +0.018 & +0.032 & +0.032 & $+0.027 \pm 0.008$ & 0.03 \\
i & 18.88 & 0.41 & +0.016 & +0.033 & +0.033 & $+0.027 \pm 0.008$ & 0.03 \\
z & 18.60 & 0.41 & +0.019 & +0.036 & +0.035 & $+0.030 \pm 0.008$ & 0.03 \\
Z & 18.61 & 0.41 & +0.006 & +0.039 & +0.076 & $+0.040 \pm 0.035$ & 0.08 \\
Y & 18.38 & 0.42 & NA & NA & NA & NA & 0.10 \\
J & 18.16 & 0.43 & +0.004 & +0.028 & +0.037 & $+0.023 \pm 0.017$ & 0.04 \\
H & 17.84 & 0.44 & -0.015 & -0.050 & +0.010 & $-0.018 \pm 0.030$ & 0.05 \\
K & 17.69 & 0.44 & -0.010 & +0.026 & +0.018 & $+0.011 \pm 0.019$ & 0.03 \\ \hline
\end{tabular}
\end{table*}

\subsection{$u-K_s$ photometry errors}
Critical to any SED fitting algorithm will be the derivation of robust
errors for each of our galaxies in each band. Here we derive the
errors from consideration of: the zero-point error ($\sigma_{\rm ZP}$),
the random sky error ($\sigma_{\rm SkyRan}$), the systematic sky error
($\sigma_{\rm SkySys}$), and the object shot noise ($\sigma_{\rm
  Shot}$). The first of these is quoted in Table~\ref{tab:zperror},
the other three can be given by:
\begin{eqnarray}
\sigma_{\rm SkyRan} = \sqrt{N_{\rm Pix}} \sigma_{\rm Sky} \\ 
\sigma_{\rm SkySys} = N_{\rm Pix} \frac{\sigma_{\rm Sky}}{\sqrt{N_{\rm Aper}}} \\
\sigma_{\rm Shot} = \sqrt{\frac{I_{\rm Obj}}{\gamma}} 
\end{eqnarray}
Where $\sigma_{\rm Sky}$ is the sky noise give in Table~\ref{tab:sky}
(Col.4), $N_{\rm Pix}$ is the number of pixels in the object aperture
(given by $\pi R_{\rm KRON}^2 A_{\rm IMAGE} B_{\rm IMAGE}$ in terms of
Source Extractor output parameters), and $N_{\rm Aper}$ is the number
of pixels used in the aperture in which the local background was
measured (i.e., $\pi R_{\rm KRON}^2 (A_{\rm IMAGE}+32) (B_{\rm
  IMAGE}+32)-N_{\rm pix}$) and $\gamma$ is the gain. Of these only the
gain is uncertain as during the stacking and renormalising of the data
the gain is modified from its original value by varying amounts (see
for example the distribution of multipliers in
Fig.~\ref{fig:viking_cuts}). However, as the vast majority of our
galaxies are relatively low signal-to-noise detections the sky errors
swamp the object shot noise errors and hence we elect to omit the
object shot noise component in our final error analysis.

\begin{figure*}

\psfig{file=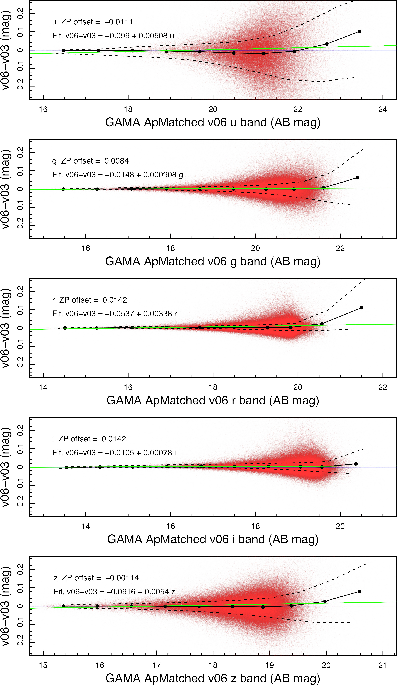,width=\columnwidth}
\psfig{file=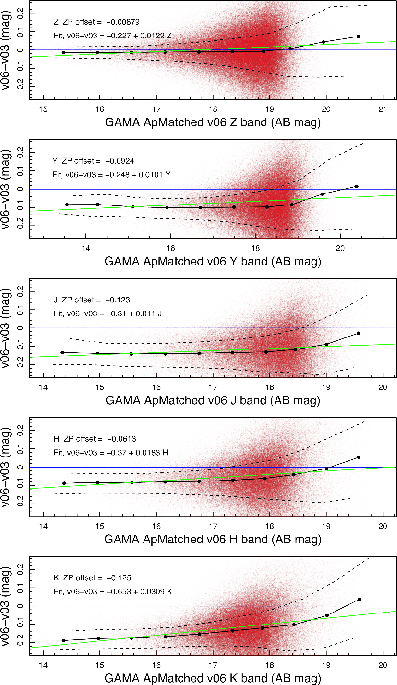,width=\columnwidth}

\caption{Comparison between GAMA ApMatchedv03 and GAMA ApMatchedv06
  (i.e., PDR) $ugrizZYJHK_s$ galaxy photometry. \label{fig:gals_opt}
  Filter transformations as indicated in the
  text. \label{fig:gals_nir}}
\end{figure*}

\subsection{Comparison to earlier GAMA photometry \label{sec:linearity}}
\begin{figure*}

\centerline{\psfig{file=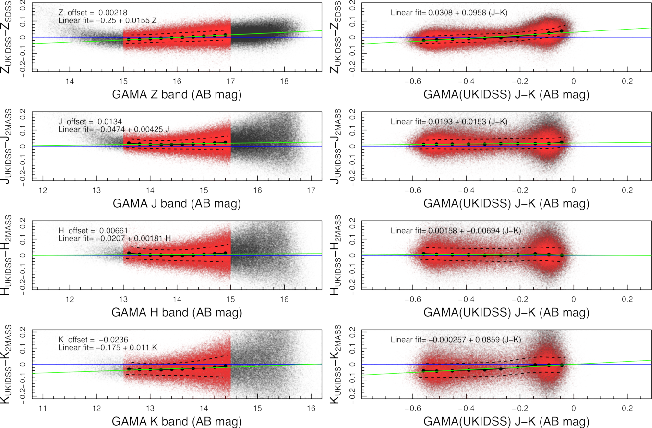,width=\textwidth}}

\caption{Comparison between 2MASS stellar photometry versus UKIDSS (IOTA) for 420k stars. \label{fig:stars2}}
\end{figure*}

Finally we compare our revised SDSS+VIKING photometry to our earlier
SDSS+UKIDSS photometry in Fig.~\ref{fig:gals_opt}. In this
implementation of IOTA the only difference in the $ugriz$ bands is the
move from a global background estimation (fixed value across the
background subtracted frame) to a local background estimation. The
impact appears minimal with zero-point offsets less than $\pm
0.015$. However in the $YJHK_s$ bands we notice significant offsets
between the UKIDSS and VIKING flux measurements. The reasons for this
are subtle and while they have not been exhaustively pursued we
believe are most likely due a hidden linearity issue in the UKIDSS
pipeline. In Fig.~\ref{fig:stars2} we show our flux measurements from
our UKIDSS data for 420k SDSS selected stars for which we have 2MASS
photometry. The agreement is once again good, however in all cases
there are significant gradients in the data and significantly stronger
than those we saw in the VIKING data
(r.f. Fig.~\ref{fig:stars_nir}). Extrapolating the linear fits to the
flux and colour regions where the majority of our galaxies lie we
infer the level of offsets seen in Fig.~\ref{fig:gals_nir}. The
implication is that there {\it may} be a linearity issue with the
UKIDSS calibration. Note that as Hill et al.~(2010) has shown our
in-house UKIDSS photometry agrees extremely well with that provided
from the UKIDSS archive. We do not explore this issue further but, as
a number of earlier GAMA papers are based on UKIDSS photometry, we
include the UKIDSS SWarps in the public release, while cautioning
against their use.

\subsection{Optical motivated far-IR photometry \label{sec:hatlas}}
To derive our far-IR photometry for every GAMA target we implement an
optically-motivated approach (also referred to as
forced-photometry). This technique closely follows the approach
developed by Bourne et al.~(2012) for the Herschel-ATLAS team and
which has been used to obtain SPIRE photometry at the location of
known optical sources. The method adopts as its starting point the
$r$-band apertures determined from our optically motivated source
finding described earlier and uses the following parameter set for the
apertures: right ascension, declination, major axis, minor axis and
position angle. For each far-IR band the aperture defined by these
parameters is combined with the appropriate PSF for each of the five
bands (supplied by the Herschel-ATLAS team). The resulting 2D
distribution therefore consists of a flat pedestal (within the
originally defined aperture region) with edges which decline as if
from the peak of the normal PSF. This soft-edge aperture can be
imagined as a 2D {\it mesa}-like distribution function which can now
be convolved with the data at the appropriate astrometric location. In
the event of two {\it mesas} overlapping the flux is shared according
to the ratio of the respective {\it mesa} functions at that pixel
location, i.e., the flux is distributed using PSF and aperture
information only and ignoring the intensity of the central
pixel. Enhancements of this methodology are under development (Wright
et al.~2015) and will include consideration of the central peak
intensity along with the inclusion of interlopers (i.e., high-z
targets), and iterations. The recovered fluxes at this moment contain
flux from the object, plus from any low-level contaminating background
objects. To assess the level of contamination we made measurements in
$\sim 30,000$ apertures of comparable sizes to our object distribution
which were allocated to regions where no known 5$\sigma$
Herschel-ATLAS detection exists nor any GAMA object. The mean
background level in these regions was found to be zero in all SPIRE
bands, as expected since the maps are made to have zero mean flux and
residual large scale emission has been removed via the nebuliser step.
In the PACS data small background values were found as shown in
Table.~\ref{tab:pacssky}. To correct for this effect the final step is
to subtract the background values for the PACS data using the {\it
  effective} aperture pixel number and factoring in shared pixels
where apertures overlap.

\begin{table}
\caption{Sky background levels per pixel as derived from blank apertures. \label{tab:pacssky}}
\begin{tabular}{c|c|c|c} \hline \hline
& \multicolumn{3}{c}{Sky background (Jy)} \\ \cline{2-4}
Band & G09 & G12 & G15 \\ \hline
100$\mu$m & 0.0002669 & 0.0001491 & 0.0002739 \\
160$\mu$m & 0.0002044 & 0.0002436 & 0.0003784 \\ \hline
\end{tabular}
\end{table}

\begin{figure}

\centerline{\psfig{file=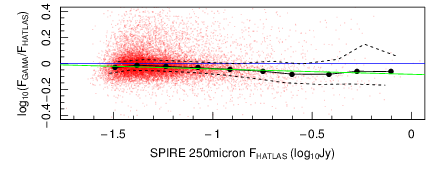,width=\columnwidth}}

\caption{A comparison between the preliminary Herschel-ATLAS 5$\sigma$
  catalogue and the optically motivated catalogue derived here in the
  SPIRE 250$\mu$m band. It should be noted that the zero-points
  calibration and entire reduction process have evolved between these
  catalogues. Note units are in Jansky as is standard in the
  far-IR. \label{fig:ben}}
\end{figure}

Figure~\ref{fig:ben} compares our aperture-matched photometry against
the Herschel-ATLAS 5$\sigma$ catalogue produced by Smith et
al.~(2012). In general the data agree reasonably well with offsets at
the levels of 0.05dex (12\%). 

There have been changes in the PACS calibration and map-making
algorithms between the generation of the values in the Smith et
al.~2012, Rigby et al.~2011 catalogues and this work, and these
changes have been substantial. Offsets at this level are consistent
with these changes. For the SPIRE data, there have been no significant
changes to the calibration or map-making process, however there are a
number of potential issues with both the catalogues being compared in
in Fig~\ref{fig:ben}. Firstly, the H-ATLAS catalogue is with a
preliminary version of the H-ATLAS release catalogue (Valiante et
al. in prep) which does not include aperture photometry for resolved
sources, explaining some of the scatter at the bright end. Secondly,
the largest optical sources, which correspond to the brightest H-ATLAS
sources are often shredded by SExtractor which leads to
inappropriately small apertures being used for the forced photometry,
and this may lead to the offset between the catalogues at the bright
end and contribute to the scatter.

In Wright et al.~(2015) we will compare our updated LAMBDAR photometry
with the final released version of the H-ATLAS catalogue (Valiante et
al.~2015) when all fluxes will be drawn from the same data pipelines
and images.

\subsection{Table matching the UV, optical/near-IR, mid-IR and far-IR catalogues}
At this stage we have a number of distinct catalogues.

~

\noindent
{\bf GalexMainv02:} This contains measurements of the FUV and NUV
fluxes which have been assembled through the use of $r$-band priors
combined with curve-of-growth analysis and is described in Liske et
al.~(2015).  We adopt the BEST photometry values. In brief the BEST
photometry is that returned by the curve-of-growth method with
automatic edge-detection when the NUV semi-major axis is greater than
20 arcsec or when the GAMA object does not have an unambiguous
counterpart. In other cases the BEST photometry is that derived from
the standard pipeline matched to the GAMA target catalogue.

~

\noindent
{\bf ApMatchedCatv06:} This contains the $u$ to $K_s$ band photometry
as described in detail in Section~\ref{sec:utoK}

~

\noindent
{\bf WISEPhotometryv02:} as described in Cluver et al.~(2014) which
outlines the detailed construction of the WISE photometry with two
exceptions. Firstly, for GAMA galaxies not resolved by WISE, standard
aperture photometry (as provided by the AllWISE Data Release) is used
instead of the profile-fit photometry (wpro). This is due to the
sensitivity of WISE when observing extended, but unresolved sources,
resulting in loss of flux in wpro values compared to standard aperture
values (see Cluver et al.~2015 for details). Secondly, the photometry
has been updated to reflect the AllWISE catalogue values. Note that
this version of the catalogue also includes the correction to the
updated W4 filter described in Brown, Jarrett \& Cluver (2014).

~

\noindent
{\bf HAtlasPhotomCatv01:} This contains the far-IR measurements as
described in Section~\ref{sec:hatlas} based on optically motivated
aperture-matched measurements incorporating contamination corrections.

~

\noindent
We use {\sc topcat} to combine these catalogues by matching on GAMA
CATAIDS (i.e., exact name matching), the FUV to $K_s$-band data are
then corrected for Galactic extinction using the $E(B-V)$ values
provided by Schlegel, Finkbeiner \& Davis (1998; GalacticExtinctionv02)
and the coefficients listed in Liske et al.~(2015).

The combined catalogue is then converted from a mixture of AB mags and
Janskys to Janskys throughout, with dummy values included when the
object has not been surveyed in that particular band. Coverage maps
may be recovered from the catalogue using the dummy values alone.

\section{Robustness checks of the PDR}
\begin{figure*}
\centerline{\psfig{file=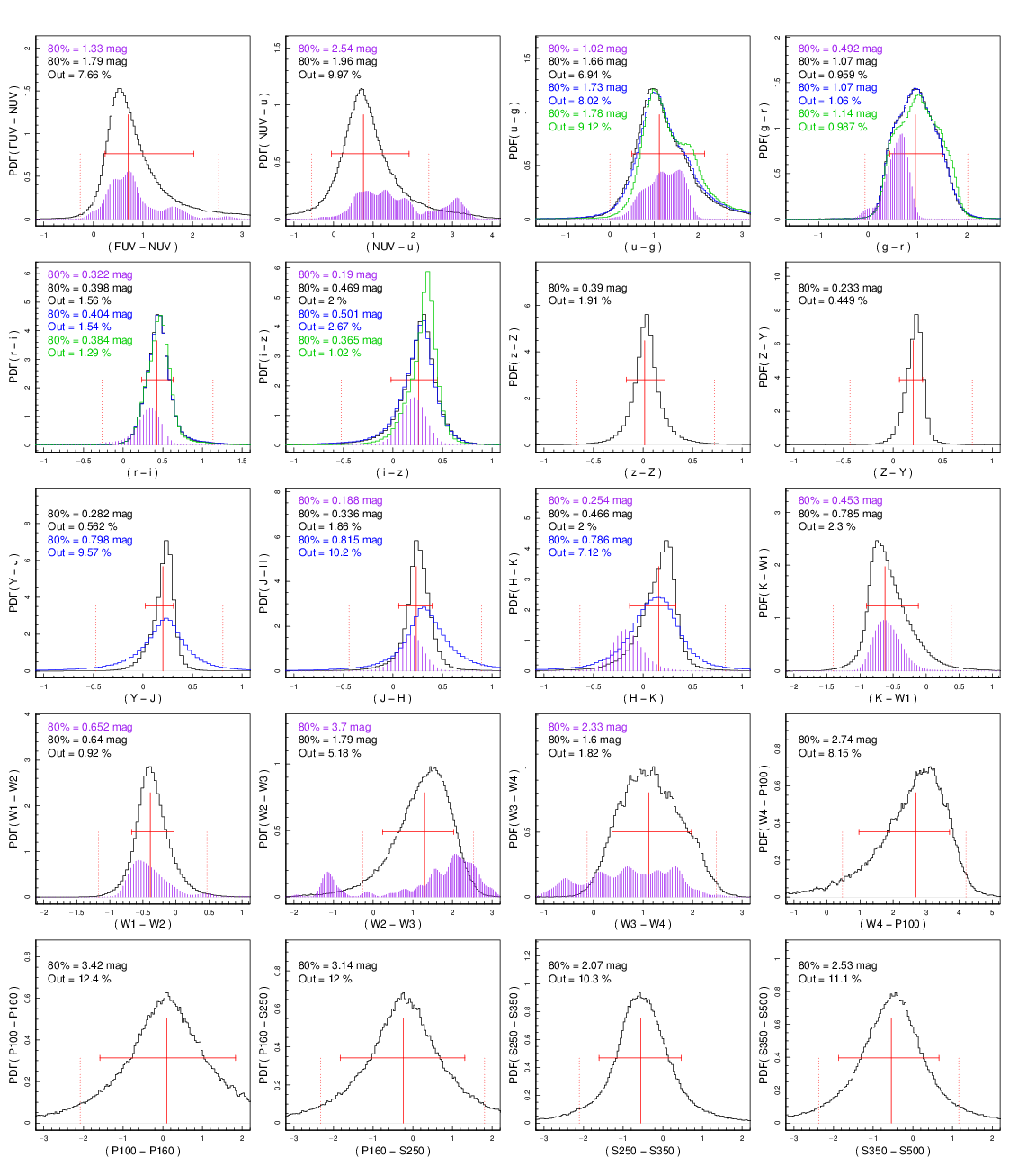,width=\textwidth}}
\caption{Each panel shows a histogram of a GAMA PDR colour histogram
  (black distribution) from two adjacent filters (as indicated,
  y-axis). The horizontal red bar indicates the 80-percentile range of
  the data with the value for GAMA PDR indicated in black in the top
  left corner. The outlier rate (indicated by Out) reports the
  percentage of the distribution which lies $>$0.5mag outside the
  80-percentile range (indicated by the vertical dotted lines). Where
  possible we overlay data from the SDSS DR7 (ModelMag colours, green
  histograms) and ApMatchedCatv03 (which uses SDSS and UKIDSS data,
  blue histograms). All distributions are extinction corrected but
  otherwise as observed. Finally the shaded histogram (purple) shows
  data from the $z=0$ sample of Brown et al.~(2014). However, note
  that the Brown sample has an ad hoc selection and essentially
  represent k-band corrected data - hence explaining the apparent
  discrepancy in $(NUV-u)$, $(u-g)$ and $(g-r)$, and $(H-K)$.
\label{fig:histograms}}
\end{figure*}

As the GAMA PDR is constructed from a variety of distinct catalogues
and pathways it is important to assess its robustness, accuracy, and
outlier rate. In earlier figures (Figs~\ref{fig:stars_opt} to
\ref{fig:stars2}) we showed direct comparisons of the magnitude
difference between two datasets. These are good for identifying
zero-point (i.e., systematic) offsets, but not particularly useful in
establishing which of the two datasets is the more robust.  Here we
examine the more informative ``colour''-plots.  The implicit
assumption is that a colour distribution arises from a combination of
the intrinsic colour spread of a galaxy population, convolved with the
measurement error in the contributing filters. A comparison of
colour-plots between two surveys, for the same sample, can provide two
important statistics: the width of the distribution, and an outlier
rate. The ``better'' quality data is the dataset with the narrowest
colour range and the lowest outlier rate (assuming the zero-points are
consistent). The colour-plot test is optimal when the intrinsic colour
spread is sub-dominant, hence should be made using adjacent
filters. In some bands, e.g., $(NUV-u)$ the intrinsic colour range is
known to be broad (e.g., Robotham \& Driver 2011), and hence the test
less conclusive. Fig.~\ref{fig:histograms} shows the full set of
colour distributions for the GAMA PDR (black histograms), the data
have been Galactic extinction corrected but not $k$-corrected, and
this is chosen to minimise the modeling dependence, particularly given
the wavelength range sampled and uncertainty in $k$-correcting certain
regimes (e.g., mid-IR).

Also shown on Fig.~\ref{fig:histograms} is the breadth of the colour
distribution derived from the 80\%-ile range (horizontal red line and
red text) and the median colour value (vertical red line).  We derive
an outlier rate (indicated by ``Out'' as a percentage on the figure),
this reports the percentage of galaxies which lie more then 0.5mag
outside the 80 percentile range. The rationale is that the 80
percentile distribution will generally capture the
intrinsic+$k$-correction spread, and a catastrophic magnitude
measurement would then be one which lies more than 0.5mag outside of
this range. One can see that the colour distributions are particularly
broad in the UV bands (as one expects given the range from
star-forming systems to inert systems with varying dust attenuation),
and in the far-IR bands (as one expects given the range of dust masses
and dust temperatures). The red optical and near-IR bands are the
narrowest (as expected given the flatness of SEDs at these
wavelengths). The outlier rates are generally highest for the poorest
resolution and lowest signal-to-noise bands (i.e., $NUV$, $FUV$, $u$
and $W2$ onwards) with outlier rates varying from 12.4 percent to 0.5
percent. Our ultimate objective within the GAMA survey is to achieve
outlier rates below 2 percent in all bands. With 10 colour
distributions at or below this level this implies we have reached this
criterion for 11 bands ($g$ to $W1$).

The facility cross-over colours, $(NUV-u)$, $(z-Z)$, $(K-W1)$,
$(W4-100\mu$m), and $(160\mu$m - 250$\mu$m), are of particular
interest as this is where mis-matches between objects might lead to
broader distributions and higher outlier rates, and there is some
indication that outlier rates do rise at these boundary points, e.g.,
the $(NUV-u)$ and $(z-Z)$ bands and $(K-W1)$ bands. In the case of the
former this may simply reflect intrinsic+$k$-correction spread.

No dataset sampling a comparable wavelength range currently exists.
However, we can compare in the optical and near-IR to the SDSS
archive, our previous catalogue (based on SDSS and UKIDSS) and into
the UV and mid-IR with the low-z templates given in Brown et
al.~(2014). These are shown where data exists as blue (GAMA
ApMatchedCatv03; SDSS+UKIDSS), green (SDSS DR7 ModelMags), and purple
histograms (Brown et al.). Note that as the SDSS and GAMA data are
essentially derived from the same base optical data and it is
the photometric measurement method which is being tested here. In
future we will be able to compare to KiDS and Subaru HSC datasets. In
comparison to our previous GAMA catalogue we can see that PDR
represents an improvement (lower breadths and lower outlier rates) in
all bands. In particular the near-IR bands are significantly improved
with the colour spread now at least two times narrower. This reflects
the greater depth of the VIKING data over the UKIDSS LAS data (see
Table~\ref{tab:sky}). In comparison to SDSS DR7 ModelMags we can see
that GAMA PDR appears to do marginally better in the $(u-g)$ and
$(g-r)$ colours, but marginally poorer in the $(r-i)$ and $(i-z)$
bands, in all cases by modest amounts.

Finally, in comparison to the colour distributions derived from the
Brown et al.~(2014) templates there are two important caveats. Firstly
the Brown data makes no attempt to be {\it statistically}
representative but rather provides an indication of the range of SEDs
seen in the nearby population for a relatively ad hoc sample. Secondly
the GAMA PDR has a median redshift of $z=0.24$ and in some bands the
k-correction will dominate over the intrinsic distribution. This is
apparent in particular in the $(NUV-u)$, $(u-g)$ and $(g-r)$ bands
where the 4000\AA~break is redshifted through. This results in
significantly broader colours in the observed GAMA PDR colour
distributions not seen in the rest-frame templates. Again this is
understandable. More puzzling
is the converse where the $(W2-W3)$ and $(W3-W4)$ distributions which
are clearly broader in the Brown et al data. This {\it may} reflect
the incompleteness within the GAMA PDR in these bands, with the bluest
objects perhaps being detected in W2 but not in W3, and hence not
represented on these plots (see Cluver et al.~2014 for full discussion
on the WISE completeness). Similarly for $(W3-W4)$. For example the
Brown et al., sample includes both Elliptical systems and very low
luminosity blue dwarf systems (e.g., Mrk 331, II Zw 96, Mrk 1490, and
UM 461) neither of which would be likely detected by WISE at $z >>
0.01$. The obvious solution is to derive ``forced-photometry'' for the
full GAMA input catalogue across all bands.

At this point we believe we have established that GAMA PDR is matching
SDSS DR7 ModelMags, a significant improvement over previous GAMA work
based on SDSS+UKIDSS LAS, but there remains some concerns regarding
higher than desired outlier rates in the lower signal-to-noise and
poorer resolution bands, and the need for a measurement at the
location of every GAMA galaxy regardless of whether there is obvious
flux or not (i.e., forced photometry). Fixing these problems is
non-trivial and requires dedicated panchromatic software, which is
currently nearing completion and will be presented in Wright et
al.~(2015).

\subsection{Composite SEDs}
Using a 35,712 core machine available at the Pawsey Supercomputing
Centre Facility (MAGNUS) we have now run the MAGPHYS spectral energy
distribution fitting code (de Cunha, Charlot \& Elbaz 2008), over the
full equatorial GAMA sample with redshifts, i.e., 197k galaxies (using
the Bruzual \& Charlot 2003 spectral synthesis model).  MAGPHYS takes
as its input, flux measurements in each band, associated errors, and
the filter bandpasses, and returns the attenuated and unattenuated SED
models from FUV to far-IR, along with a number of physical
measurements, e.g., stellar mass, star-formation rate, dust mass, dust
opacity, dust temperatures (birth-cloud and ISM) etc (for more details
please see da Cunha et al.~2008). Here we look to use MAGPHYS to
provide a simple spectral energy representation for each of our
galaxies which also has the effect of filling in the gaps where
coverage in a particular band does not exist or no detection is
measured. On a single processor MAGPHYS will typically take 10mins to
run for a single galaxy (i.e., 4yrs for our full sample), but using
MAGNUS the entire sample can be processed in less than 24hrs.

\begin{figure*}
\centerline{\psfig{file=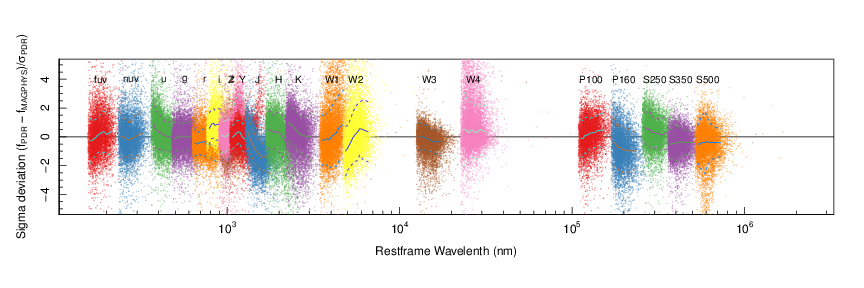,width=\textwidth}}
\caption{Sigma deviations of the model photometry compared to the
  input photometry versus rest wavelength for all data points for all
  galaxies. For each colour the median and 1$\sigma$ dispersions is
  measured and shown (grey solid and dotted lines respectively). The
  plot shows that the majority of our data lies within the quoted
  error of the MAGPHYS fits with some outliers (as also noted in
  Fig.~\ref{fig:histograms}). The offset of W4 is of some concern as
  is the bleeding of the far-IR data towards negative deviation
  values. Both of these effects are believed to be understood but
  requires the LAMBDAR software (in development). If these are flat
  and offset it implies a systematic offset, if they show trends with
  wavelength it implies a progression of a feature with
  redshift. Trends are most apparent in the near-IR where the
  modelling of the TP-AGB population is still uncertain.
  \label{fig:angusplot}}
\end{figure*}

Fig.~\ref{fig:angusplot} shows each of the $21\times 197,491$
datapoints plotted at the rest wavelength versus the number of
$\sigma$ deviations the GAMA PDR magnitude is away from the derived
MAGPHYS magnitude. The figure highlights that generally MAGPHYS
appears to be finding consistent fits across all bands with only the
W4 showing some indication of a fundamental inconsistency between the
data and the models. This has now been tracked down to the change in
the W4 filter transmission curve with our MAGPHYS run still using
the old throughput curve while the WISE PDR data uses the revised W4
transmission curve (see Brown, Jarrett \& Cluver 2014 for full
details). Future runs of MAGPHYS will use the updated curve.  The
distribution of the data points in $\sigma$ deviations (abscissa)
suggest that the measurement variations are consistent with the errors
quoted. The bleed of the far-IR data to the lower part of the figure,
is most likely due to contamination by high-z systems (as
expected). In the wavelength range, within each filter the near-IR
data shows the most fluctuations. These are likely to reflect
recurrent features in the MAGPHYS models shifting through
the various bands and suggests some uncertainty in the precise
modelling of the TP-AGB region as noted by numerous groups, e.g.,
Maraston et al.~(2006).

\begin{figure*}
\centerline{\psfig{file=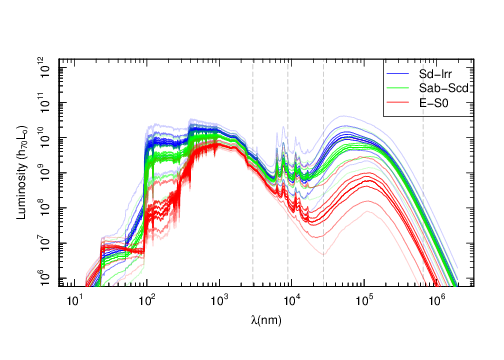,width=\textwidth}}
\caption{The panel shows the 10, 25 ,45 ,50 ,55 ,75 and 90 percentiles
  for the full GAMA PDR SEDs for the $z<0.06$ morphologically
  classified sample presented in Moffett et al.~(2015). The data are
  initially scaled to the same mass and then at each wavelength point
  the quantiles derived which then collectively trace out the
  quantiles over wavelength. The spread at each wavelength therefore
  directly reflects the spread in the mass-to-light ratio at that
  wavelength.
  \label{fig:angusplots}}
\end{figure*}

Fig.~\ref{fig:angusplots} shows the MAGPHYS SED model fits sampled by
our $z<0.06$ morphologically classified sample, separated into E/S0s
(red), Sabcs (green) or Sd/Irrs (blue). See Moffett et al.~(2015) for
details on the sample and morphological classification process. In
order to construct these plots the individual SEDs derived by MAGPHYS
have been normalised to the same stellar mass, i.e., their SEDs have
been scaled by their fractional stellar mass offset from
$10^{10}M_{*}$ (using the stellar masses derived by MAGPHYS). The
curves shown are the quantile distributions for 10, 25, 45, 50, 55, 75
and 90 percent as a function of wavelength. Note that these do not
represent individual MAGPHYS SED models, but are quantile ranges in
narrow wavelength intervals which are then linked to create the SED
quantiles - hence the SEDs show more variations than the models used
in MAGPHYS if examined in detail. The reason for this representation
is to avoid a specific calibration wavelength and SEDs that, when
calibrated into quantiles at one wavelength, cross at others.

Fig.~\ref{fig:angusplots} illustrates not only the wealth of data
provided by the GAMA PDR but a number of physical phenomena. Firstly
the spread at any wavelength point represents the mass-to-light ratio
at that wavelength. This can be seen to be narrowest in the 2 -
5$\mu$m range (as expected), indicating that this region is optimal
for single band stellar-mass estimates. However, the constant gradient
in this region implies that near-IR colours provide little further
leverage to improve the stellar-mass estimation beyond single band
measurements. Conversely the smooth variation of SED gradients in the
optical from low to high stellar-mass ratios, imply that optimal
stellar-mass estimation may arise from the combination of a single
band near-IR measurement combined with an optical colour (see also
discussion in Taylor et al.~2011). Fig.~\ref{fig:angusplots}
highlights the known strong correlations between UV flux, far-IR
emission and stellar mass-to-light ratio with all being amplified or
suppressed in Sd/Irrs or E/S0s respectively. However, that all
galaxies seem to contain some far-IR emission may be a manifestation
of the MAGPHYS code tending to maximise dust content within the bounds
as allowed by the far-IR errors. Curiously the Sabc (green) provide a
very narrow range of parameters, perhaps indicating a close coupling
between star-formation, dust production, and the mass-to-light ratio
--- arguably indicative of well-balanced self-regulated disk
formation/evolution. The greater spreads in the early (red) and later
(blue) types are perhaps indicative of the progression through various
stages of quenching (ramping down) and unstable disc formation
(overshoot) respectively. In particular Agius et al.~(2015) found an
unexpectedly high levels of dust in a significant (29\%) population of
the GAMA-E/S0 galaxies consistent with a range of E-So SEDS.  Note
also the results presented on observed correlations between the
star-formation rate, specific star-formation rate in da Cunha et
al.~(2010) and Smith et al.~(2012); see also interpretation in Hjorth
et al.~(2014). A detailed exploration of these phenomena are beyond
the scope of this paper but the potential is clear particularly in
conjunction with the existing group (Robotham et al.~2011) and large
scale structure (Alpaslan et al.~2014) catalogues.

\subsubsection{Inspection of individual objects}
\begin{figure*}

\vspace{-1.8cm}

\psfig{file=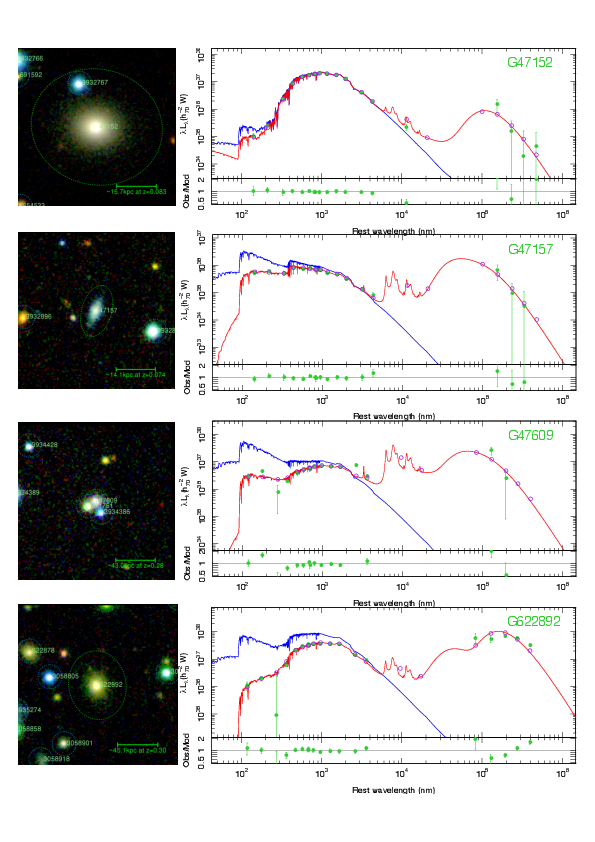,width=\textwidth}

\vspace{-1.8cm}

\caption{Four example galaxies from the GAMA PDR showing, in each
  case, the $Hig$ image with the apertures overlaid, and the 21-band
  photometry with the attenuated (red) and unattenuated (blue) MAGPHYS
  fits to the data. Data shown in green represent the measured
  photometry plotted at rest wavelength while the purple circles show
  the integrated flux measured from the attenuated MAGPHYS fit
  through the appropriate filter. The lower portion of the main panels
  shows the residuals, i.e., the ratio of the observed flux to the
  model flux. Errors include an arbitrary 10 percent error added in
  quadrature intended to incorporate some allowance for zero-point
  offsets between facilities. Note that the lowest image is a well
  known lens system reported by Negrello et al.~(2010). All images are
  derived from the GAMA $\Psi$ online cutout tool
  http://gama-psi.icrar.org/ \label{fig:magphys}}
\end{figure*}

We explore individual SEDs for one hundred systems randomly selected
(IDs 47500-47609). Approximately 5-10 percent are found to have one or
more significant outlier(s) in the photometry but otherwise good
MAGPHYS fits are found for all 100 systems. In approximately 50
percent of cases the far-IR photometry is essentially missing (due to
the shallowness of the far-IR data), hence flux and/or redshift cuts
are advisable depending on the science investigation to be
conducted. Fig.~\ref{fig:magphys} shows four example galaxies which
include a nearby bright system (G47152, $z=0.082$), a nearby faint
system (G47157, $z=0.074$), a higher redshift crowded system (G47609,
$z=0.282$), and a known far-IR lens system (G622892, $z=0.300$;
Negrello et al.~2010; recently shown to exhibit a spectacular Einstein
ring, the very high far-IR flux is evident). The panels on the left
show the combined $giH$ colour image from a combination of VIKING and
SDSS data. Overlaid (green dotted lines) are apertures for the main
object and nearby systems in our bright catalogue. The right panels
show the 21-band measured photometry (green data and errorbars) in
units of total energy output ($\lambda L_{\lambda}$ in units of
h$_{70}$ W) at the filter pivot-wavelength divided by $(1+z)$ (i.e.,
rest wavelength). The red and blue lines show the attenuated and
unattenuated SEDs from the preliminary MAGPHYS fits.  Purple circles
show the flux from the attenuated SED curve integrated within the
filter bandpasses given in Fig.~\ref{fig:filters}. The lower portion
of the panel shows the residuals expressed as the ratio of the
observed flux to the measured flux. Included in the error budget is a
10 percent flux component added in quadrature to mitigate small
systematic zero-point offsets at facility boundaries. Comparable plots
for all 221k systems with redshifts are provided via the GAMA $\Psi$
online cutout tool (http://gama-psi.icrar.org/)

The four galaxies are each well sampled by the GAMA PDR data which
collectively map out the two key peaks in the energy output due to
starlight and dust reprocessing of starlight. The four systems also
show varying degrees of dust attenuation with the observed fluxes
requiring minimal, significant and extreme corrections to recover the
unattenuated fluxes. In all cases the residuals are well behaved,
given the errors. Exploring the GAMA PDR more generally, the SED data
appear robust with catastrophic failures in one band typically at the
levels indicated in Fig.~\ref{fig:histograms}, i.e., 10 percent in
poorly-resolved bands to 1 percent in well-resolved bands. Obvious
issues which arise following inspection of several hundred SEDs are:
incorrect apertures, data artifacts (i.e., poor quality regions),
nearby bright stars (diffraction spikes and blocking), crowding, and
confusion.

\section{The energy output of the Universe at $z<0.2$ from FUV to far-IR \label{sec:energy}}
We conclude this paper with a brief look at the integrated energy
output of the low redshift galaxy population, i.e., the Cosmic
Spectral Energy Distribution (CSED), its recent evolution, and the
implied integrated photon escape fraction. The CSED represents the
energy output of a cosmologically representative volume, in essence an
inventory of the photons recently generated, as opposed to those
passing through but formed earlier. It can be reported both pre- and
post- attenuation by the dust content of the galaxy population, both
are interesting. The pre-attenuated CSED informs us of the photons
being created from (primarily) nucleosynthesis processes (in the
current epoch), while the post-attenuated CSED informs us of the
photons entering into the IGM. The sum of the two must equal (energy
conservation), but the wavelength distribution will differ as dust
re-processes the emergent photons from short (UV and optical) to long
(mainly far-IR) wavelengths. The combination of the two can be used to
determine the integrated photon escape fraction. By integrated we
imply over a representative galaxy population, and averaged over
representative viewing angles. Both of these factors are important and
make the integrated photon escape fraction useful for converting
observed FUV and NUV fluxes to robust star-formation rates.  The work
follows earlier measurements of the CSED reported in Driver et
al.~(2008) and Driver et al.~(2012). However, the methodology here is
very different and for the first time includes mid- and far-IR data in
a fully consistent analysis. In our earlier studies we determined
luminosity distributions in each band independently and then fitted
across these values to determine the CSED. Here we stack the
individual MAGPHYS SEDs fits derived earlier, as representative
fitting functions (see Fig.~\ref{fig:magphys}).

Potentially, as we have a MAGPHYS fit for every galaxy we could use
them to derive fluxes in data gaps and use the full sample. However,
given the critical importance of the far-IR dust constraint we elect
to use just the common region with full 21-band coverage (see
Figs.~A\ref{fig:g09} to~A\ref{fig:g23}). This combined region
constitutes an area of 63 percent of the full area or 113~deg$^2$ and
contains 138k objects with secure redshifts in the range $0.02 < z <
0.5$ (trimmed to exclude stars and high-z AGN). To explore any
evolution of the CSED, we divide our sample into three redshift
intervals: $0.02 < z <0.08$, $0.08 < z < 0.14$ and $0.14 < z< 0.2$
which correspond roughly to lookback times of 0.8~Gyr, 1.5~Gyr and
2.25~Gyr respectively.  The volumes sampled are: $4.9 \times 10^5$,
$2.1 \times 10^6$ \& $4.6 \times 10^6$ h$_{70}^{-3}$ Mpc$^{3}$
respectively (factoring in our reduced coverage). Within each redshift
range we use the $z_{\rm max}$ values reported in Taylor et al.~(2011)
to derive a weight as not all galaxies would be visible across the
selected redshift range. Galaxies with $z_{\rm max}$ values above the
redshift range have weights set to unity, and values with $z_{\rm
  max}$ below this range have weights set to zero. Otherwise weights
are set to the inverse of the fraction of the volume sampled. A cap is
placed ensuring no weight exceeds a value of 10, this ensures a single
lone system just fortuitously within the redshift range cannot
dominate the final outcome by being massively amplified.  Within each
redshift range we now simply sum the energy $\times$ weight (i.e.,
$\sum W_i \lambda L^i_{\lambda}$) for the galaxies within our
selection to arrive out the CSED for that volume.

These raw derived CSEDs require one final correction to accommodate
for the loss of lower luminosity systems in the higher redshift
bins. To determine the correction factor we repeat the summation but
with a stellar mass cut imposed on all three samples
($10^{10}$M$_{\odot}$), this is sufficiently low to be sampling the
dominant contribution to the CSED but not so low as to suffer total
incompleteness (i.e., that which is not corrected for by our
weights). For each volume interval we obtain unrestricted to
mass-restricted CSED ratios of: 1.68, 1.48 and 1.29 for the low, mid,
and high redshift samples respectively. If all three samples were
complete this ratio would be constant, hence this changing ratio
encodes the loss of the lower luminosity systems in the higher
redshift bins, and can therefore be used to provide an appropriate
correction. This is achieved by scaling the final CSED curves by
factors of 1.00, 1.14 (i.e., 1.68/1.48), and 1.30 (i.e., 1.68/1.29)
respectively. In effect we are using the CSED shape from the
unrestricted samples but normalising using the restricted samples and
this is analogous to the normalisations typically used in estimating
luminosity and mass functions. This implicitly assumes the following:
that the low redshift sample is itself complete (hence requiring no
scaling), and that the ratio of energy emerging from systems above and
below $10^{10}$M$_{\odot}$ is approximately constant. The first of
these is relatively secure: GAMA is a deep survey and at $z \sim 0.08$
is mass complete to $10^9$M$_{\odot}$ (see Lange et al.~2015), below
which there is very little contribution to the luminosity density (see
Driver~1999), or stellar mass density (see Moffett et al.~2015). In
the second case we understand low-mass systems are preferentially
star-forming and may have evolved more rapidly over recent times
compared to the more massive systems. However, as the correction
factors are relatively modest (14 percent and 30 percent) the shape
and renormalisation is unlikely to be dramatically changed, but we
acknowledge may be biased low. This can only be quantified through
deeper studies (see for exampled the planned WAVES survey; Driver et
al.~2015).

\begin{figure*}
\centerline{\psfig{file=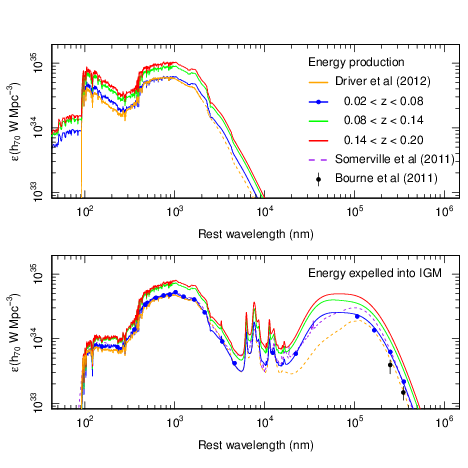,width=\textwidth}}
\caption{The energy originating (i.e., unattenuated, top), and
  emanating (i.e., attenuated following dust reprocessing, lower) at
  intervals equivalent to 0.75, 1.5 and 2.25~Gyr lookback time. The
  data are normalised to the energy output per Mpc$^3$ for
  $H_o=70$km/s/Mpc. The data show clear trends in the evolution of the
  total energy output over this timeline. \label{fig:energy}}
\end{figure*}

Fig.~\ref{fig:energy} shows the resulting energy outputs for the
renormalised unattenuated CSED (upper) and the renormalised attenuated
CSED (lower) with the redshift ranges represented by colour as
indicated. Also shown is our earlier estimate derived from GAMA via
luminosity function fitting (orange line). It is worth re-iterating
that the Driver et al.~(2012) CSED measurement is based on luminosity
density measurements from $FUV$ to $K_s$ combined with an adopted
integrated photon escape fraction (Driver et al.~2008) to infer the
mid and far-IR portion, i.e., the mid and far-IR from Driver et
al. (2012) is a prediction, and hence shown as a dotted line beyond
$2.1\mu$m. Also shown is the prediction from semi-analytic modelling
by Somerville et al.~(2012). Both curves, corrected to $H_o=70$
km/s/Mpc, follow the new low-z CSED very well in the optical and start
to diverge in the mid and far-IR bands where previous empirical data
have been lacking. The Somerville curve, in particular traces the
low-redshift bin extremely well with the previous Driver data
significantly under-predicting the far-IR emission. The CSEDs
presented here represent a major advance constituting the first
consistent measurement of the post- attenuated CSED from a single
sample spanning from the $FUV$ to far-IR. Hence, while cosmic (sample)
variance (CV) may scale the respective CSEDs in overall density, it
will not modify the shape of the distribution (unless there are
extreme hidden clustering factors). Using the formula in Driver \&
Robotham (2010, see also the online calculator
http://cosmocalc.icrar.org/) we derive the CV, for our three redshift
ranges to be: 18 percent, 12 percent and 10 percent respectively, with increasing
redshift and note that despite the extent of the GAMA PDR the dominant
error remains the CV.

To first order the CSEDs are therefore all consistent with each other
and the previous values.  The CSEDs follow the expected progression
towards higher energy output towards higher redshifts, despite the
potential uncertainty in overall normalisation from CV.  In particular
the far-IR increases noticeably faster than the optical. This result
is independent of cosmic variance and noted previously within the
far-IR community, see for example Dunne et al.~(2011) who infer a
significant increase in dust mass (towards high redshift) over the past
5~Gyr. In detail we can report that the Universe is in energy decline,
having dropped from a total energy production of $(2.5 \pm 0.2) \times
10^{35}$ h$_{70}$ W Mpc$^{-3}$ at 2.25~Gyr ago, to $(2.2 \pm 0.2)
\times 10^{35}$ h$_{70}$ W Mpc$^{-3}$ 1.5~Gyr ago and $(1.5 \pm 0.3)
\times 10^{35}$ h$_{70}$ W Mpc$^{-3}$ 0.75~Gyr ago. This decline is
significant despite the CV uncertainty, and in line with our
understanding of the evolution in the cosmic star-formation history
(e.g., Hopkins \& Beacom 2006), which shows a decline of a factor of
approximately $\times 1.5$ over this time-frame. The lowest redshift
bin value is also consistent with the $(1.26 \pm 0.09) \times 10^{35}$
h$_{70}$ W Mpc$^{-3}$ for $z<0.1$ as reported in Driver et al.~(2012)
and the orange and blue curves on Fig.~\ref{fig:energy} do show
consistency over the optical regime. The majority of the energy
difference is derived from the far-IR where previous data was lacking
and the CSED in this region estimated. As an aside it is also worth
noting that the near-IR data show some slight disagreement, and this
is consistent with our finding that VIKING fluxes are typically
brighter than UKIDSS LAS fluxes (see
Section~\ref{sec:linearity}). Finally, shown on Fig.~\ref{fig:energy}
are the SPIRE data points derived by the Herschel-ATLAS team from the
initial data release which sampled all of G09, G15 and half of G12
(Bourne et al.~2012). These data are low compared to our new estimate,
however the SPIRE calibration and data reduction have evolved quite
substantially since these data-points were derived.

Integrating these distributions reveals some interesting numbers.
From the unattenuated CSED we find that 50 percent of the energy production
in the nearby Universe is at wavelengths in the range
0.01---0.64$\mu$m (i.e., UV/optical). In the post-attenuated
``observed'' CSED, 50 percent of the energy emerges at wavelengths shorter
than 1.7$\mu$m. Splitting at 10$\mu$m we find that 65 percent of the energy
produced (via stellar emission) is released into the IGM at shorter
wavelengths and 35 percent at longer wavelengths. Despite dust contributing
a very small proportion of a galaxy's mass (typically $<1$ percent; Driver
et al.~2008), its impact on the energy output is dramatic with
significant potential consequences for optically based flux and size
measurements (see for example Pastrav et al.~2013).

\begin{figure}
\centerline{\psfig{file=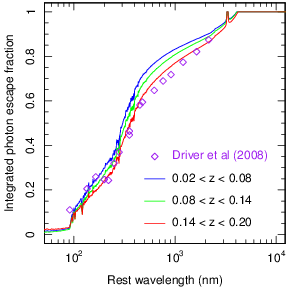,width=\columnwidth}}
\caption{The photon escape fraction integrated over all viewing angles
  and derived from Fig.~\ref{fig:energy}. Also shown is the escape
  fraction reported by Driver et al.~(2008) from the Millennium Galaxy
  Catalogue within the redshift range $0.0 < z < 0.18$. The data show
  a clear progression towards lower escape fractions as redshift
  increases. \label{fig:escape}}
\end{figure}

Dividing the pre- and post-attenuated CSEDs yields the integrated
photon escape fraction (IPEF). This is a particularly interesting
distribution as it encapsulates the impact of dust in a simple and
general way. Fig.~\ref{fig:escape} shows the IPEF in the three
redshift bins. Also shown (purple diamonds) is the IPEF derived in
Driver et al.~(2008) for the Millennium Galaxy Catalogue (Liske et
al.~2003). Clearly apparent is a trend towards lower photon escape
fractions towards modestly higher redshift, demonstrating the impact
of dust evolution in our perception of even low redshift systems
($z<0.2$). It is important to note that this result is resilient to CV
as we are comparing the ratio of the pre- and post- attenuated CSEDs
and hence the normalisation cancels out. The variation appears smooth
with redshift and significant, with the escape fraction in the FUV
changing from 18 percent in the higher redshift bin to 23 percent in
the lower z bin. The corresponding change in the NUV is 27 percent to
34 percent respectively. If this trend continues the implication is
that at even intermediate redshifts the UV photon escape fraction may
be significantly lower than the values typically adopted when deriving
star-formation rates from UV fluxes.

The above work should be considered as preliminary, but indicative of
the potential of the GAMA PDR to explore the energy outputs of
galaxies and galaxy population at low redshift. The analysis at
present also includes a number of important caveats which we are
looking to address in the near future. The first is out-standing
issues related to the GAMA photometry: aperture robustness, the need
for a uniform forced photometry across all bands, and improved
management of the variable spatial resolutions and signal-to-noise
limits. The second is whether MAGPHYS provides truly unbiased fits,
particularly in the far-IR where our data quality is lowest and where,
as the errors grow, over-fitting of the far-IR fluxes cannot be ruled
out. Verifying that MAGPHYS is unbiased will be important as our data
quality undoubtedly declines with redshift. One should also bear in
mind that the interpretations above are very much based on the
assumptions embedded in the MAGPHYS code, in due course it will be
important to explore a range of assumptions and to undertake critical
comparisons against fully radiative-transfer codes for well-resolved
systems. The third and arguably most fundamental issue relates back to
the integrity of the GAMA input catalogue and in particular its
reliance on the fairly shallow SDSS imaging and the potential for
missing extended low surface brightness systems. The GAMA regions are
currently being surveyed by the VST KiDS team and will also be
surveyed as part of the HSC Wide survey, both datasets can be used to
improve our input catalogue particularly for extended low surface
brightness systems.

\section{Summary}
We have brought together a number of diverse datasets from three space
missions (GALEX, WISE and Herschel), and two ground-based facilities
(SDSS and VISTA) to produce the GAMA Panchromatic Data Release. The
individual dataframes have been astrometrically and photometrically
matched, and then SWarped into single images for each of the three
equatorial GAMA regions covering 80~deg$^2$ each (i.e., slightly
larger than the nominal 60~deg$^2$ region covered by the
spectroscopy). Weight maps are provided indicating the number of
frames contributing to each region and files containing the names of
the individual frames used. In the $u-K_s$ bands we provide both
native resolution data and data degraded to a common 2$''$ full-width
half maximum. The SWarped images along with a cutout tool (GAMA
$\Psi$) for extracting subregions are available at:

~

{\bf http://gama-psi.icrar.org/}

~

Note that the Herschel data is currently proprietary but will be made
available following the Herschel-ATLAS final data release.

GAMA $\Psi$ also provides additional functionality to create
colour images on the fly from the combination of any of the 21-bands,
individual fits downloads, object and aperture overlays and
is maintained by the International Centre for Radio Astronomy Research
(ICRAR) Data Intensive Astronomy (DIA) unit. Queries and comments on
GAMA $\Psi$ should be sent to simon.driver@uwa.edu.au.

We describe the construction of the $u-K_s$ aperture-matched
photometry following the method described in Hill et al.~(2011) and
compare to our previous measurements. In addition we compare our
near-IR photometry for 400k stars extracted from 2MASS and confirm
that our zero-points are robust to within a tenth of a magnitude.

The VIKING-2MASS and UKIDSS-2MASS zero-points are shown to be robust
however VIKING-UKIDSS photometry do not agree. This apparent
tautological inconsistency is most likely explained by a linearity
issue as the VIKING and UKIDSS data are compared at a significantly
fainter magnitude range.  Some evidence is seen for a linearity issue
in the UKIDSS-2MASS comparison, which when extrapolated to fainter
magnitudes does appear to explain the offsets seen between VIKING and
UKIDSS.

We also describe our method for deriving far-IR measurements using
optical priors where we use our $r$-band apertures convolved with the
appropriate instrument/filter PSF to measure the far-IR flux at the
location of every object.

We then combine our $u-K_s$ and $100-500\mu$m fluxes with GALEX and
WISE data derived by the MPIK and UCT/UWC groups led by RJT and MC
using exact name ID matching. This results in a final catalogue of
over 221k objects from the G09, G12, G15 and G23 regions, 63 percent of which
have complete coverage in all 21 bands.

Finally as a demonstration of this dataset we derive the total flux
originating and emanating from various volumes from FUV to far-IR at
redshifts indicative of 0.8, 1.5 and 2.25~Gyr lookback time. We see
evidence for evolution over this period consistent with the decline of
star-formation traced by the cosmic star-formation history and
consistent with the reported evolution of the far-IR luminosity
functions.

In future papers we will further improve our flux extraction method
using optically motivated priors in all bands, leading to more
consistent errors, and explore the physical properties derived from
SED fitting codes for various populations and sub-populations.

\section*{Acknowledgments}
LD and SJM acknowledge support from the European Advanced Investigator
grant cosmicism. SB acknowledges support from an ARC Future Fellowship
(FT140101166). NB acknowledges support from EC FP7 SPACE project
ASTRODEEP (Ref.No: 312725).  E. Ibar acknowledges funding from
CONICYT/FONDECYT postdoctoral project N$^\circ$:3130504.  MALL
acknowledges support from UNAM through PAPIIT project IA101315.  PN
acknowledges the support of the Royal Society through the award of a
University Research Fellowship, the European Research Council, through
receipt of a Starting Grant (DEGAS-259586) and support of the Science
and Technology Facilities Council (ST/L00075X/1).

We acknowledge the use of data products from the NASA operated GALEX
space mission. We also acknowledge the following institutions and
agencies for their financial contributions towards the reactivation
and operations of the GALEX satellite. This has allowed us to complete
NUV observations of the G23 and Herschel ATLAS SGP regions: The
Australian-Astronomical Observatory (AAO), the Australian Research
Council (ARC), the International Centre for Radio Astronomy Research
(ICRAR), the University of Western Australia, the University of
Sydney, the University of Canterbury, Max-Plank Institute f\"ur
KernPhysik (MPIK), the University of Queensland, the University of
Edinburgh, Durham University, the European Southern Observatory (ESO),
the University of Central Lancashire, Liverpool John Moore University,
National Aeronautics and Space Administration (NASA), Universit\'e
Paris Sud, University of California Irvine, Instituto Nazionale Di
Astrofisica (INAF), and the University of Hertfordshire.

Funding for the SDSS and SDSS-II has been provided by the Alfred
P. Sloan Foundation, the Participating Institutions, the National
Science Foundation, the U.S. Department of Energy, the National
Aeronautics and Space Administration, the Japanese Monbukagakusho, the
Max Planck Society, and the Higher Education Funding Council for
England. The SDSS Web Site is http://www.sdss.org/.

The SDSS is managed by the Astrophysical Research Consortium for the
Participating Institutions. The Participating Institutions are the
American Museum of Natural History, Astrophysical Institute Potsdam,
University of Basel, University of Cambridge, Case Western Reserve
University, University of Chicago, Drexel University, Fermilab, the
Institute for Advanced Study, the Japan Participation Group, Johns
Hopkins University, the Joint Institute for Nuclear Astrophysics, the
Kavli Institute for Particle Astrophysics and Cosmology, the Korean
Scientist Group, the Chinese Academy of Sciences (LAMOST), Los Alamos
National Laboratory, the Max-Planck-Institute for Astronomy (MPIA),
the Max-Planck-Institute for Astrophysics (MPA), New Mexico State
University, Ohio State University, University of Pittsburgh,
University of Portsmouth, Princeton University, the United States
Naval Observatory, and the University of Washington.

The VIKING survey is based on observations with ESO Telescopes at the
La Silla Paranal Observatory under the programme ID 179.A-2004.

This publication makes use of data products from the Wide-field
Infrared Survey Explorer, which is a joint project of the University
of California, Los Angeles, and the Jet Propulsion
Laboratory/California Institute of Technology, funded by the National
Aeronautics and Space Administration.

The Herschel-ATLAS is a project with Herschel, which is an ESA space
observatory with science instruments provided by European-led
Principal Investigator consortia and with important participation from
NASA. The H-ATLAS website is http://www.h-atlas.org/.

GAMA is a joint European-Australasian project based around a
spectroscopic campaign using the Anglo-Australian Telescope. The GAMA
input catalogue is based on data taken from the Sloan Digital Sky
Survey and the UKIRT Infrared Deep Sky Survey. Complementary imaging
of the GAMA regions is being obtained by a number of independent
survey programs including GALEX MIS, VST KiDS, VISTA VIKING, WISE,
Herschel-ATLAS, GMRT and ASKAP providing UV to radio coverage. GAMA is
funded by the STFC (UK), the ARC (Australia), the AAO, and the
participating institutions. The GAMA website is
http://www.gama-survey.org/ . 

This work is supported by resources provided by the Pawsey
Supercomputing Centre with funding from the Australian Government and
the Government of Western Australia.

\section*{References}

\reference Abazajian K.N., et al., 2009, ApJS, 182, 543 
\reference Agius N., et al., 2015, MNRAS, in press (arXiv:1505.06947) 
\reference Alpaslan M., et al., 2014, 438, 177
\reference Andrae E., 2014, PhD Thesis, MPIfK
\reference Arnaboldi M., et al., 2007, Messenger, 127, 28 
\reference Baldry, I.K., et al., 2010, MNRAS, 404, 86 
\reference Barnes D.G., et al., 2001, MNRAS, 322, 486
\reference Bate, M, Bonnell, I.A., Bromm, V., 2003, MNRAS, 339, 577
\reference Bertin E., 2010, Astrophysics Source Code Library, record ascl.1010.068 
\reference Bertin E., 2011, ASPC, 442, 435 
\reference Bourne N., et al., 2012, MNRAS, 421, 3027 
\reference Brown M.J.I., Moustakas J., Smith J.D.T., da Cunha E., Jarrett T.H., Imanishi M., Armus L., Brandl B.R., \& Peek J.E.G., 2014, ApJS, 212, 18
\reference Brown M.J.I., Jarrett T.H., Cluver M.E., 2014, PASA, 31, 49 
\reference Bruzual G., Charlot S., 2003, MNRAS, 344, 1000
\reference Calzetti D., Armus L., Bohlin R.C., Kinney A.L., Koornneef J., Storichi-Bergmann T., et al., 2000, ApJ, 533, 682
\reference Cluver M., et al., 2014, ApJ, 782, 90
\reference Cluver M., et al., 2015, MNRAS, in preparation
\reference da Cunha E., Charlot, S., Elbaz D., 2008, MNRAS, 388, 1595 
\reference da Cunha E., Eminian C., Charlot, S, Blaizot J., 2010, MNRAS, 403, 1894
\reference Dalton G., et al., 2006, SPIE, 6269
\reference de Jong J.T.A., Verdoes Kleijn, G.A., Konrad, K.H., Valentijn E.A., 2013 Experimental Astronomy, 35, 25
\reference Drinkwater M.J., et al., 2010, MNRAS, 401, 1429
\reference Driver S.P., 1999, ApJ, 526, 69
\reference Driver S.P., et al., 2007, MNRAS, 379, 1022
\reference Driver S.P., et al., 2009, A\&G, 50, 12
\reference Driver S.P., et al., 2008, ApJ, 678, 101 
\reference Driver S.P., et al., 2011, MNRAS, 413, 971
\reference Driver S.P., et al., 2012, MNRAS, 427, 3244 
\reference Driver S.P., Davies L.J., Meyer M., Power C., Robotham A.S.G., Baldry I.K., Liske J., Norberg P., 2015, ASSP, in press (arXiv: 1507.00676) 
\reference Dom\'inguez A., et al., 2011, MNRAS, 410, 2556 
\reference Dunne L., et al., 2011, MNRAS, 417, 1510
\reference Eales S. et al., 2010, PASP, 122, 499 
\reference Edge A., Sutherland W., Kuijken K., Driver S.P., McMahon R., Eales S., Emerson J.P., 2013, Messenger, 154, 32
\reference Fontanot F., Monaco P., Cristiano S., Tozzi P., 2006m MNRAS, 373, 1173
\reference Giavalisco M., et al., 2004, ApJ, 600, 93
\reference Griffin M.J., et al., 2010, A\&A, 518, 232 
\reference Grogin N.A., et al., 2011, ApJS, 197, 35
\reference Hill D., et al., 2010, MNRAS, 404, 1215 
\reference Hill, D., et al., 2011, MNRAS, 412, 765 
\reference Hambly N.C., et al., 2001, MNRAS, 326, 1279
\reference Hjorth J., Gall C., Michalowski M.J., 2014, MNRAS, 782, 23
\reference Hopkins A.M., et al., 2013, MNRAS, 430, 2047
\reference Hopkins A.M., Beacom J.F., 2006, ApJ, 651, 142
\reference Hopkins P.F., Hernquist L, Cos T.J, Di Matteo T, Robertson B, Springel, V, 2006, ApJS, 163, 1
\reference Ibar E., et al., 2010, MNRAS, 409, 38
\reference Jarrett T.H., et al., 2012, AJ, 144, 68
\reference Jarrett T.H., et al., 2013, AJ, 145, 6
\reference Kelvin L., et al., 2012, MNRAS, 421, 1007
\reference Kelvin L., et al., 2014, MNRAS, 444, 1647
\reference Kennicutt R.C., et al., 2003, PASP. 115, 928
\reference Keres D., Katz N, Weinberg D, Dav\'e R., 2005, MNRAS, 363, 2
\reference Koekemoer A.M., et al., 2011, ApJS, 197, 36
\reference Komatsu E., et al., 2011, ApJS, 192, 18 
\reference Lacey C., Cole S., 1993, MNRAS, 262, 627
\reference Lange R., et al., 2015, MNRAS, 447, 2603
\reference Lewis J.R., Irwin M., Bunclark P., 2010, ASPC, 434, 91
\reference Liske J., et al., 2003, MNRAS, 344, 307 
\reference Liske J., et al., 2015, MNRAS, 452, 2087 
\reference Maddox S.J., et al., 2015, MNRAS, in preparation
\reference Maraston C., Daddi E., Renzini A., Cimatti A., Dickinson M., Papovich C., Pasquali A., Pirzkal N., 2006, ApJ, 652, 82
\reference Martin C., et al., 2005, ApJ, 619, 1 
\reference Masci F.J., 2013, online only, arXiv:1301.2718
\reference Masci F.J., Fowler J.W., 2009, ASPC, 411, 67
\reference McKee C.F., Ostriker E.C., 2007, ARA\&A, 45, 565
\reference Moffett A., et al., 2015, MNRAS, submitted
\reference Morrissey P., et al., 2007, ApJS, 173, 682
\reference Munoz-Mateos J.C., et al., ApJS, in press (arXiv:1505.03534)
\reference Negrello et al., 2010, Science, 330, 800
\reference Pascale E., et al., 2011, MNRAS, 415, 911
\reference Pastrav B., Popescu C.C., Tuffs R.J., Sansom A.E., 2013, A\&A, 557, 137
\reference Pilbratt G.L., et al., 2010, A\&A, 518, 1
\reference Poglitsch A., et al., 2010, A\&A, 518, 2
\reference Popescu C.C., Tuffs R.J., 2002, MNRAS, 335, 41
\reference Rigby E.E., et al., 2011, MNRAS, 415, 2336
\reference Robotham A.S.G., Driver S.P., 2011, MNRAS, 413, 2570
\reference Robotham A.S.G., et al., 2011, MNRAS, 416, 2640
\reference Robotham A.S.G., et al., 2010, PASA, 27, 76 
\reference Rowlands K., et al., 2012, MNRAS, 419, 2545 
\reference Schlegel D.J., Finkbeiner D.P., Davis M., 1998, ApJ, 500, 525
\reference Schoenberner D., 1983, ApJ, 272, 708
\reference Scoville, N., et al., 2007, ApJS, 172, 1 
\reference Sheth K., et al., 2010, PASP, 122, 1397
\reference Shu F., Adams F.C., Lizano S., 1987, ARA\&A, 25, 23
\reference Skrutskie M.F., et al., 2006, AJ, 131, 1163
\reference Smith D., et al., 2011, MNRAS, 416, 857
\reference Smith D., et al., 2012, MNRAS, 427, 703
\reference Somerville R., Gilmore, R.C., Primack J.R., Dom\'inquez A., 2012, MNRAS, 423, 1992
\reference Soto M., et al., 2013, A\&A, 552, 101
\reference Soifer B.T, Neugenbauer G., Hoick J.R., 1987, ARA\&A, 25, 187 
\reference Sutherland R., et al., 2015, A\&A, 575, 25
\reference Tuffs R.J., Popescu C.C., V\"olk H.J., Kylafis N.D., Dopita M.A., 2004, A\&A, 419, 821
\reference Taylor, N., et al., 2011, MNRAS, 418, 1587
\reference Toomre A., \& Toomre J., 1972, ApJ, 178, 623
\reference Tinsley B., 1980, Fund. Cosmic Physics, 5, 287
\reference Valiante E., et al., 2015, MNRAS, in preparation
\reference Veilleux S., Cecil G., Bland-Hawthorn J., 2005, ARA\&A, 43, 769
\reference White R.L., Becker R.H., Helfand D.J., Gregg M.D., 1997, ApJ, 475, 479
\reference White S.D.M., Rees M.J., 1978, MNRAS, 183, 341
\reference Wright A., et al., 2015, MNRAS, in preparation
\reference Wright E.L., et al., 2010, AJ, 140,1868 
\reference York D., et al., 2000, AJ, 120, 1579

\appendix

\begin{figure*} 

\psfig{file=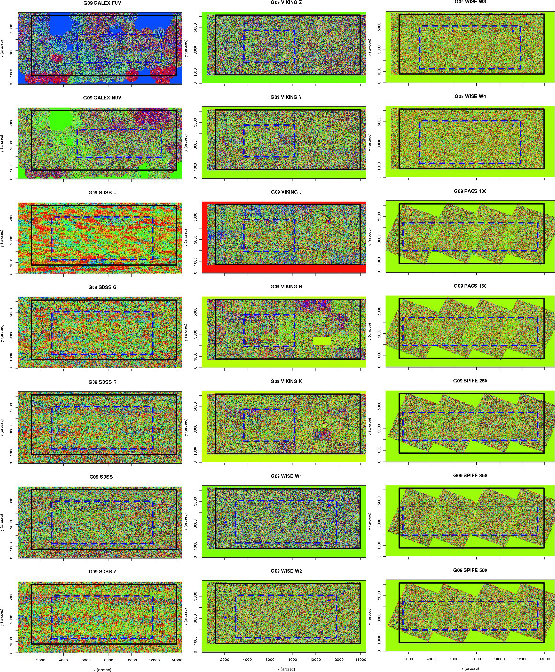,width=\textwidth}

\caption{Background uniformity and coverage in the G09 region for
  GALEX, SDSS, VIKING, WISE and Herschel data. The black box denotes
  the GAMA spectroscopic survey region and the blue box the region
  from which the background statistics were derived. We use the
  MOGRIFY package to display the low-res data frames very close to the
  sky level. Astronomical objects will not be visible, however the
  frames highlight the coverage, missing regions, and the integrity of
  the large scale sky structure. \label{fig:g09}}

\end{figure*}

\begin{figure*} 

\psfig{file=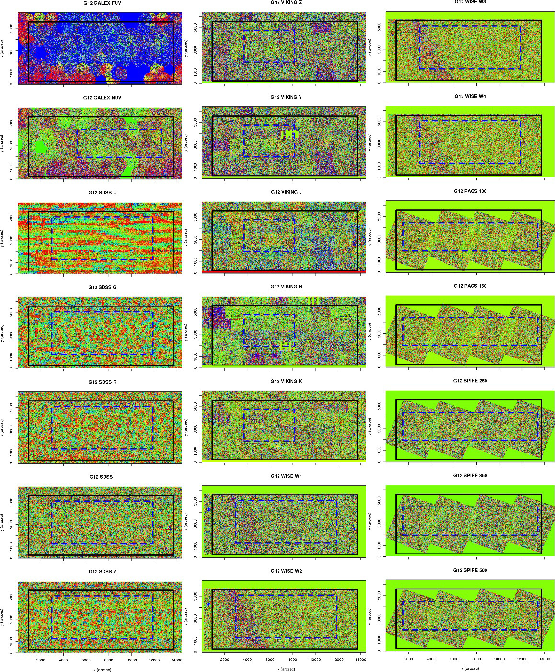,width=\textwidth}

\caption{Background uniformity and coverage in the G12 region for GALEX, SDSS, VIKING, WISE and Herschel data. \label{fig:g12}}

\end{figure*}

\begin{figure*} 

\psfig{file=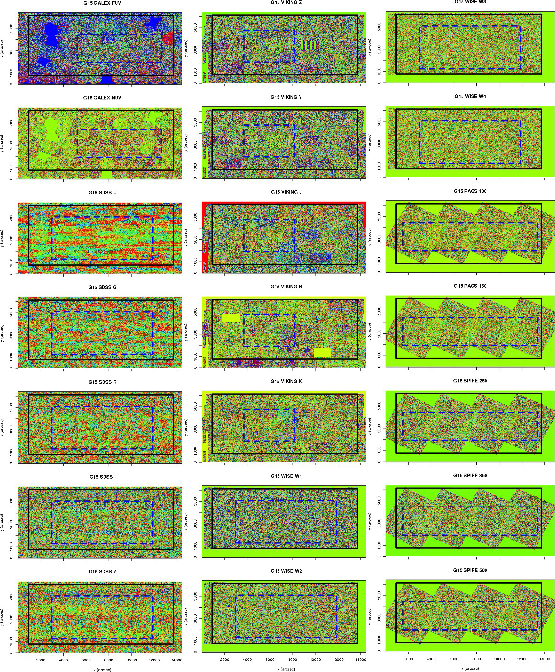,width=\textwidth}

\caption{Background uniformity and coverage in the G15 region for GALEX, SDSS, VIKING, WISE and Herschel data. \label{fig:g15}}

\end{figure*}

\begin{figure*} 

\psfig{file=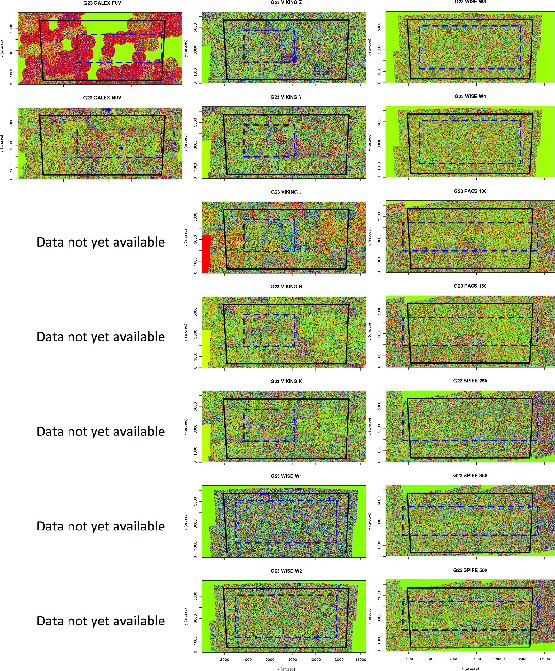,width=\textwidth}

\caption{Background uniformity and coverage in the G23 region for GALEX, VIKING, WISE and Herschel data. \label{fig:g23}}

\end{figure*}

\section{GAMA PDR coverage}
Figs.~\ref{fig:g09} to \ref{fig:g23} shows the GAMA PDR coverage in
each of the 21 bands, all data is available for download via:
http://gama-psi.icrar.org/

\label{lastpage}

\end{document}